\documentclass[10pt,a4paper,oneside]{article}
\usepackage[cp850]{inputenc}
\usepackage{amsmath,amsfonts,latexsym,amssymb}
\usepackage{epic,eepic}
\usepackage{epsfig}
\usepackage{graphicx}

\DeclareMathOperator{\sign}{sign}
\DeclareMathOperator{\tr}{tr}

\DeclareMathOperator{\str}{str}

\DeclareMathOperator{\diag}{diag}

\DeclareMathOperator{\arccosh}{arccosh}

\DeclareMathOperator*{\Simiq}{\simeq}
\DeclareMathOperator*{\Simi}{\sim}

\newcommand{\vect}[1]{{\mathbf #1}}
\newcommand{\vectgr}[1]{{\boldsymbol#1}}    
\newcommand{\Frac}[2]{\displaystyle\frac{#1}{#2}}
\newcommand{\smc}[1]{\text{\sc{#1}}}

\title{\vspace{-3\baselineskip}%
              Optimal Fluctuations and Tail States of non-Hermitian
              Operators}

\author{\small F. M. Marchetti${}^{1,2}$ and B. D. Simons${}^{1}$ \\
       \small ${}^{1}$ \emph{Cavendish Laboratory, Madingley Road,
       Cambridge CB3 \ OHE, UK }\\
       \small ${}^{2}$ \emph{Scuola Normale Superiore, Piazza dei
       Cavalieri 7, 56126 Pisa, Italy}}

\date{\small October 22, 2001}

\begin{document}

\maketitle

\begin{abstract}
\emph{A statistical field theory is developed to explore the density
of states and spatial profile of `tail states' at the edge of the 
spectral support of a general class of disordered non-Hermitian
operators. These states, which are identified with symmetry broken,
instanton field configurations of the theory, are closely related to
localized sub-gap states recently identified in disordered
superconductors. By focusing separately on the problems of a quantum
particle propagating in a random imaginary scalar potential, and a
random imaginary vector potential, we discuss the methodology of our
approach and the universality of the results. Finally, we address
potential physical applications of our findings.}
\end{abstract}

\section{Introduction}
\label{sec:intro}
Over recent years the statistical properties of stochastic
non-Hermitian operators have come under intense
scrutiny~\cite{sommers,haake,hatano_nelson,efetov1,brouwer,fyodorov,janik,feinberg_zee,chalker_wang1,brezin_zee,mudry_simons,chalker_mehlig,hastings,janik2,izyumov_simons1,izyumov_simons2,yurkevich,kolesnikov_efetov,mudry_brouwer}.
Operators of this kind appear in a number of physical applications ranging
from random classical dynamics, and statistical physics, to phase
breaking and relaxation in quantum dynamics. For example, a
straightforward mapping shows the statistical mechanics of a repulsive
polymer chain can be described in terms of the quantum dynamics of a
particle subject to a random imaginary scalar
potential~\cite{edwards,kleinert}. Similarly, the statistical
mechanics of flux lines in a type-II superconductor pinned by a
background of impurities can be described as the quantum evolution of
a particle in a disordered environment subject to an imaginary vector
potential~\cite{hatano_nelson}. Finally, the diffusion of a classical 
particle advected by a random velocity field is described by a linear
\emph{non-Hermitian} operator, the ``Passive Scalar''
equation~\cite{chalker_wang1}. More generally, the dynamics of various
classical systems can be expressed in terms of random Fokker-Planck
operators (for a review see, e.g.,
Refs.~\cite{bouchaud_georges,isichenko}). 

Non-Hermitian operators exhibit several qualitatively new phenomena
which discriminate their behaviour from those of their random
Hermitian counterparts. Firstly, while the eigenvalues of Hermitian
operators are real, the spectrum of non-Hermitian operators is
generically complex, bound by some region of support in the
complex plane. The second striking  distinction concerns the
localization properties of the eigenfunctions. It is well established
that a weak random impurity potential brings about the localization of
all eigenstates of a Hermitian operator in dimensions of two and
below~\cite{abrahams_anderson,gorkov_larkin}. By contrast, a constant
imaginary vector potential is sufficient to delocalize states of the
disordered system even in one-dimension! This phenomenon, which was 
reported and explained by Hatano and Nelson~\cite{hatano_nelson}, finds 
extension to higher dimensions~\cite{kolesnikov_efetov}.

Beginning with early work on random matrix
ensembles~\cite{ginibre,girko}, a variety of techniques have been
developed to study the statistical properties of stochastic
non-Hermitian 
operators~\cite{sommers,haake,hatano_nelson,efetov1,brouwer,fyodorov,janik,feinberg_zee,chalker_wang1,brezin_zee,mudry_simons,chalker_mehlig,hastings,janik2,izyumov_simons1,izyumov_simons2,yurkevich,kolesnikov_efetov,mudry_brouwer}.
Some of the techniques have been based on the random matrix 
theory~\cite{ginibre,girko,sommers,haake,fyodorov,janik,feinberg_zee,janik2},
while others have involved the development of perturbative
schemes, such as the self-consistent Born approximation in the
diagrammatic analysis (e.g., Ref.~\cite{chalker_wang1}). In this paper
we will be concerned with adapting a third approach which involves the
refinement of field theoretic techniques which have proved to be so
useful in the description of random Hermitian operators.

Recently, attention has been directed towards the study of 
`tail states' which persist at the edge of the spectral
support of the non-Hermitian system. Indeed,
there are already indications~\cite{balagurov_vaks,samokhin,shnerb}
that an important role can be played by those parts of the spectrum
which are populated by rare states. Now, in stochastic \emph{Hermitian}
systems a small tail of states accumulate below the band edge tightly
bound to rare or `optimal fluctuations' of the random
potential~\cite{lifshitz,zittartz_langer,halperin_lax,gredeskul}.
These so-called `Lifshitz tail' states 
are typically smooth, node-less, localized functions which inhabit
regions where the potential is particularly shallow and smooth. By
contrast, the character of the complex spectrum in non-Hermitian
operators allows for the existence of tail states that occupy the
entire region which bounds the support of the spectrum. 

The difference is not incidental: tail states in the non-Hermitian
system can exhibit features characteristic of the quasi-classical states
within the bulk spectrum. To emphasize the point, consider the spectrum of,
say, a two-dimensional free quantum particle subject to a complex scalar
potential which involves both real \emph{and} imaginary components,
\begin{equation}
  \hat{H} = \Frac{\hat{\vect{p}}^2}{2m} + i V(\vect{r}) + W (\vect{r})
  \; ,
\label{eq:hamil}
\end{equation}
where $\hat{\bf p}=-i\hbar\nabla$. If the imaginary potential $V$ is
absent, a region of localized tail states accumulate below the band
edge ($\epsilon<0$) due to optimal configurations of the potential. At
energies $\epsilon$ greatly in excess of the inverse scattering time
$\hbar/\tau$, bulk states are only `weakly localized' with a
localization length $\xi_{\text{loc}}$ greatly in excess of scattering
mean-free path $\ell$ and wavelength $\lambda$. This is the
quasi-classical regime where mechanisms of quantum interference
strongly affect the spectral and transport properties of the
system~\cite{abrahams_anderson,gorkov_larkin}.

\begin{figure}
\begin{center}
\includegraphics[width=0.6\linewidth,angle=0]{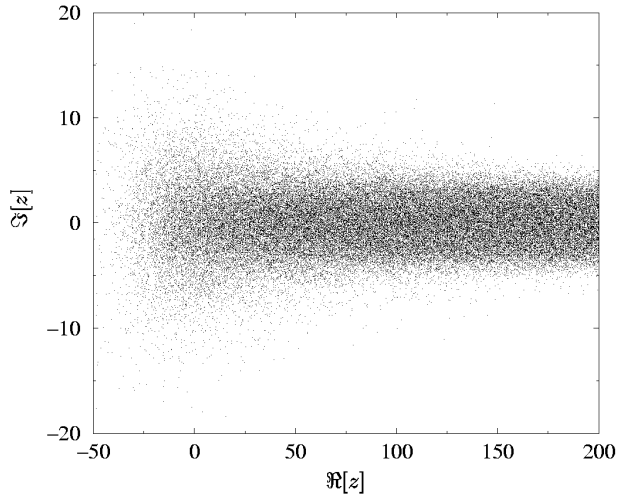}
\end{center}
\caption{\small
        Complex eigenvalue spectrum for several realizations of the 
        random non-Hermitian Hamiltonian~(\ref{eq:hamil}), where the 
        potentials $V$ and $W$ are both drawn from a Gaussian 
        $\delta$-correlated random impurity distribution. Here we have 
        included a lattice of $29 \times 29$ $\vect{q}$-points in the
        momentum space.}
\label{fig:specs}
\end{figure}

On this background, let us now suppose that a small imaginary
component of the potential is restored. Being non-Hermitian, the
eigenvalues migrate into the complex plane occupying a finite 
region of support centered on the real axis (see
Fig.~\ref{fig:specs}). At the level of the mean-field analysis
(detailed below), complex eigenvalues are confined to a sharp region
of support in the complex plane. However, as with Lifshitz states,
optimal fluctuations of the random potential(s) generate states which
lie outside the region of support (see Fig.~\ref{fig:specs}). What is
the nature of these states? Are they localized and, if
so, over what length scale? Here, in contrast to the Hermitian system, tail
states must accommodate fast fluctuations of the wave function at the
scale of the wavelength of the Hermitian system.

To our knowledge, the problem of tail states in the non-Hermitian
system was first considered in a recent work by Izyumov and 
Simons~\cite{izyumov_simons2}. Refining an
approach developed in Refs.~\cite{halperin_lax,gredeskul} to explore
tail states in Hermitian systems, the authors introduced a
non-perturbative scheme to investigate states at the edge of the
support in non-Hermitian systems. This instanton technique revealed
the existence of tail states at the edge which are localized on a
length scale greatly in excess of the wavelength --- tail states in
the non-Hermitian system are quasi-classical in nature. Building on 
the work of Efetov~\cite{efetov}, the aim of the present paper is to 
develop a supersymmetric field theory to explore the spatial profile 
of the states in the tail region, and to determine the complex 
density of states (DoS) in the vicinity of
the band edge. In doing so, we will reveal a close correspondence
between tail states in the non-Hermitian problem and sub-gap states in
weakly disordered superconductors. Moreover, constraints within the present
scheme reveals the universal character of tail states in the
non-Hermitian system showing that, after a suitable rescaling, the
profile of the DoS tail depends only on dimensionality and separation
from the support. In particular, in contrast to Lifshitz states in the
Hermitian system, properties of the tail states are universal, independent 
of the nature of the random distribution.

To illustrate the generality of our approach we will consider two examples 
which, according to the classification scheme defined below, belong to
different fundamental symmetry. Firstly, following our discussion
above, we will investigate the properties of tail states in a quantum 
system involving the propagation of a particle in a random \emph{scalar}
potential involving both real \emph{and} imaginary 
components~(\ref{eq:hamil}). Secondly, we will investigate the nature of
tails states in a quantum system where the particle is subject to a 
random real scalar potential and a random imaginary \emph{vector} potential,
\begin{equation}
  \hat{H} = \Frac{[ \hat{\vect{p}} + i \vect{h} (\vect{r})]^2}{2m} +
  W(\vect{r})  \; .
\label{eq:hamiv}
\end{equation}
As well as the Hatano-Nelson system~\cite{hatano_nelson}, where
$\vect{h}=\text{const}$, \eqref{eq:hamiv} has been studied in the
context of random classical dynamics involving classical diffusion in
the background of a quenched random velocity field. Identifying
$1/2m$ with the classical diffusion constant $D$, and
$\vect{h}/2m$ with a random velocity field $\vect{v}$, a redefinition
of the random potential $W(\vect{r})$ obtains the classical operator
\begin{displaymath}
  \hat{H} = - D \nabla^2 + (\nabla \cdot \vect{v}) + \vect{v} \cdot
  \nabla + \bar{W} (\vect{r})\; .
\end{displaymath}

In the quasi-classical approximation, (i.e. setting $z$ as the complex 
energy, where $\tau \Re [z]\gg \hbar$) our goal will be to show that the 
long-range, low-energy properties of the random systems are described by a 
supersymmetric non-linear $\sigma$-model. At the level of mean-field, the
spectrum of both non-Hermitian operators will be shown to be bound to
a region of support inside the complex plane. By taking into account
non-perturbative symmetry breaking instanton field configurations of 
the action, we will reveal the existence of tail states which extend 
beyond the region of support predicted by the mean-field. We will
determine the localization properties and profile of the DoS revealing
an intrinsic universality of the present scheme. 

The paper is organized as follows: after introducing a general scheme 
to classify linear non-Hermitian operators on the basis of their
fundamental symmetries, in section~\ref{sec:field} we formulate a 
statistical field theory to describe the spectral properties of the 
random imaginary scalar potential Hamiltonian. Within this approach, 
we show that the large scale properties of the spectral support are 
determined by the saddle-point or mean-field properties of the theory.
Motivated by a parallel description of the disordered superconducting 
system, in section~\ref{sec:state} we show that optimal fluctuations 
of the random impurity potential induce localized tail states which 
extend beyond the region of support. Within this approach, we determine
analytical expressions for the spatial extent of the localized states 
in the tail region, and determine the scaling of the complex DoS as a 
function of energy. Generalizing the scheme to the consideration of the 
imaginary vector potential Hamiltonian in section~\ref{sec:vecto}, we 
emphasize the universality of the present scheme. Results from both 
sections are compared with numerical simulations. Finally, in 
section~\ref{sec:concl} we conclude our discussion.

\section{Field Theory of non-Hermitian Operators}
\label{sec:field}
The aim of this section is to prepare a statistical field theory of
the weakly disordered non-Hermitian system. To be concrete, in the
first instance, we will focus on the problem of the imaginary scalar 
potential~\eqref{eq:hamil}. Later, we will see that the field theoretic
scheme is easily modified to accommodate the constant vector potential
perturbation.
However, before embarking on this program, it is useful to first
contrast general properties of complex linear non-Hermitian operators
with those of their Hermitian counterparts. As well as helping to
clarify the idiosyncrasies of the non-Hermitian system, we will expose
a general scheme in which non-Hermitian operators can be classified 
according to their fundamental symmetries.

\subsection{Background: Symmetry Classification}
\label{sec:backg}
A general non-Hermitian operator $\hat{H}$ is specified by left and 
right eigenfunctions,
\begin{align*}
  {\hat H}|R_i\rangle &= z_i |R_i\rangle & \langle L_i|{\hat H} &=
  \langle L_i|z_i \; ,
\end{align*}
where $\{z_i\}$ denotes the set of complex eigenvalue. Although not a 
generic feature of non-Hermitian operators, throughout we will rely on the 
assumption that the eigenfunctions form a complete basis set. This is 
the case if there are no repeated eigenvalues, which will be true for
a generic random operator such as those studied here. Furthermore we
will take the eigenstates to be orthonormal, 
\begin{align}
  \langle L_i | R_j \rangle &= \delta_{ij} & \sum_i |R_i\rangle
  \langle L_i| &=\mathbb{I} \; . 
\label{eq:ortho}
\end{align}

Now as with Hermitian operators, spectral (and localization)
properties of the non-Hermitian Hamiltonian can be expressed through
the complex Green function
\begin{equation*}
  \hat{g}(z)=\Frac{1}{z-\hat{H}} \; ,
\end{equation*}
where $z=x+iy$ denotes the complex argument. Inserting the resolution of 
identity~\eqref{eq:ortho}, one obtains the spectral decomposition,
\begin{equation}
  \hat{g}(z)=\sum_i | R_i\rangle\Frac{1}{z - z_i} \langle L_i| \; .
\label{eq:cgree}
\end{equation}
Using the analytical properties of the resolvent, it is
straightforward to show that the density of complex eigenvalues
\begin{displaymath}
\nu (z) = \Frac{1}{L^d} \tr \delta (z - \hat{H}) = \Frac{1}{\pi L^d}
\lim_{\eta \to 0} \sum_i \Frac{\eta^2}{(|z - z_i|^2 + \eta^2)^2} \;,
\end{displaymath}
is expressible in terms of the complex Green function through the
identity~\eqref{eq:cgree}:
\begin{equation*}
  \nu (z) = \Frac{1}{\pi L^d} \partial_{z^\ast} \tr \hat{g} (z) \; .
\end{equation*}
The complex Green function $\hat{g} (z)$ is, therefore, a non-analytic
function everywhere in the complex $z$-plane in which the
corresponding eigenvalues density is non-vanishing. 

Previous studies have shown that techniques based on standard
diagrammatic perturbation theory account only for contributions to
$\hat{g}(z)$ which are \emph{analytic} in $z$~\cite{sommers}. To
account for all contributions, it is convenient to follow the now
standard route~\cite{girko,feinberg_zee,chalker_wang1} of expressing
the Green function through an auxiliary Hamiltonian which is
explicitly Hermitian. This is achieved by constructing a matrix
Hamiltonian, $\hat{\mathcal{H}}$, with the following $2\times 2$ block
structure
\begin{equation*}
\hat{\mathcal{H}} = 
\begin{pmatrix}
0 & \hat{H} - z \\
\hat{H}^\dag - z^\ast & 0
\end{pmatrix} \; .
\end{equation*}
In this representation, the Green function of the non-Hermitian
operator is expressed as the off-diagonal element of the matrix Green 
function~\cite{chalker_wang1},
\begin{equation*}
\hat{g}(z)=\lim_{\eta\to 0}\hat{G}_{21}(\eta , z) \; ,
\end{equation*}
where $\hat{G} (\eta , z) = (i\eta-\hat{\mathcal{H}})^{-1}$.

This `method of Hermitization' affords a convenient way of classifying
the symmetries of a non-Hermitian Hamiltonian according to the
fundamental symmetries of its Hermitian counterpart. The utility of
this classification will become manifest presently in the construction 
of the statistical field theory of the non-Hermitian system. As an 
example, let us consider the Hamiltonian involving an imaginary scalar 
potential~\eqref{eq:hamil}. Applied to this problem, the Hermitization
procedure leads to the matrix Hamiltonian
\begin{equation*}
\hat{\mathcal{H}} = \left( \hat{\zeta}_{\hat{\vect{p}}}-x+W \right)
\sigma_1 + \left(y - V\right)\sigma_2 \; ,
\end{equation*}
where the Pauli matrices $\sigma_i$ act in the $2\times 2$ sub-space, 
and $\hat{\zeta}_{\hat{\vect{p}}}=\hat{\vect{p}}^2/2m$. 

In fact, the matrix structure of $\hat{\mathcal{H}}$ bares much in
common with the Gor'kov or Bogoliu\-bov-de Gennes
quasi-particle Hamiltonian of a weakly disordered 
superconductor~\cite{izyumov_simons1}. To
foster this analogy let us implement the canonical transformation
\begin{equation}
\hat{\mathcal{H}} \mapsto \hat{\mathcal{H}}_\Gamma = \Gamma^{-1}
\hat{\mathcal{H}} \Gamma = \left( \hat{\zeta}_{\hat{\vect{p}}} - x + W
\right)\sigma_3 + \left(y - V\right) \sigma_2 \; , 
\label{eq:gprime}
\end{equation}
where $\Gamma=\exp[- i\pi\sigma_2/4]$. Associating the matrix structure 
with a `particle/hole' space, the operator
$\hat{\mathcal{H}}_\Gamma$ can be identified as a Gor'kov Hamiltonian
for a disordered superconductor. Here $x$ plays the role of the
chemical potential, while $\hat{\zeta}_{\hat{\vect{p}}}+W$ denotes the
bare Hamiltonian, and $(y-V)$ represents the (in this case real,
random) order parameter. However, taking the analogy further, the
energy argument in the Gor'kov Green function corresponds, in this
case, to the infinitesimal $i\eta$ of the matrix Green function
$\hat{G}^{-1}_{\Gamma} (\eta , z)$ --- evidently, we are interested in
the zero energy quasi-particle states of the disordered, time-reversal
invariant, superconductor with a random real order parameter.

In its Hermitian form, the Hamiltonian above can be classified
according to its fundamental symmetries. According to the
classification introduced by Zirnbauer~\cite{zirnbauer}, 
the matrix construction places a general non-Hermitian Hamiltonian 
in the chiral unitary symmetry class denoted AIII\footnote{\
  \label{note1} 
  The notation adopted by Zirnbauer is motivated by that introduced by
  Cartan to classify the $10+1$ symmetric spaces.}, a general element
of which has the form 
\begin{displaymath}
\begin{pmatrix}
0 & Z \\
Z^\dag & 0
\end{pmatrix} \; ,
\end{displaymath}
where $Z$ is an arbitrary complex matrix. However, within this group, 
there are four subclasses of higher symmetry: class CI (the
time-reversal invariant superconductor) where matrices $Z$ are complex
symmetric; class BDI (chiral orthogonal) where the elements of $Z$ are 
arbitrary and real; class CII where $Z$ has an underlying symplectic 
structure; and class DIII where $Z$ is complex antisymmetric. As 
will be clear from the forthcoming analysis, the statistical field
theory describing the non-Hermitian spectral properties has soft modes
associated with each universality class. From our discussion above,
the imaginary scalar potential is accommodated in the symmetry class
CI.

As the second example, let us consider the non-Hermitian
Hamiltonian~\eqref{eq:hamiv} involving the random imaginary vector 
potential. In this case, the Hermitization procedure leads to the matrix
Hamiltonian 
\begin{equation}
  \hat{\mathcal{H}}_\Gamma = \left(\Frac{\hat{\vect{p}}^2 -
  \vect{h}^2}{2m} + W - x\right) \sigma_3 + \left(y -
  \Frac{\hat{\vect{p}} \cdot \vect{h} + \vect{h} \cdot
  \hat{\vect{p}}}{2m}\right) \sigma_2 \; .
\label{eq:vdoub}
\end{equation}
As a result, we find that the system belongs to a different class 
depending on the value of the imaginary component of the energy argument
$y$. Along the real axis (i.e. for $y=0$) $\hat{\mathcal{H}}$ is
real and therefore belongs to the symmetry class BDI. Away from the real
axis the Hamiltonian becomes complex and the symmetry is lowered to class
AIII. Later we will discuss the physical manifestations of the discrete
symmetries in the two limits.

\subsection{Generating Functional}
\label{sec:gener}
With this background, let us now turn to the construction of a field
theory to describe statistical correlations of the non-Hermitian
model. Motivated by the correspondence outlined above, the analysis
will parallel previous investigations of the weakly disordered
superconductor~\cite{altland_simons,bundschuh,simons_altland}. The
starting point is the generating functional for the single-particle
Green function. Adopting throughout the convention $\hbar = c = 1$, 
the latter is expressed as a field integral involving two independent 
four component supervector fields $\psi(\vect{r})$ and
$\bar{\psi}(\vect{r})$ with elements having both
commuting/anti-commuting and `particle/hole' indices:
\begin{displaymath}
  \mathcal{Z}[j] = \int D (\bar{\psi} , \psi) \exp \left[i \int d
  \vect{r} \; \bar{\psi} \left( i \eta - \hat{\mathcal{H}}\right) \psi
  + \int d \vect{r} \left(\bar{\psi} j +\bar{j} \psi\right)\right] \;
  .
\end{displaymath}
Elements of the Green function can be generated from the source terms
$j(\vect{r})$ and $\bar{j}(\vect{r})$. For example, from the
generating function one can obtain the DoS according to the identity:
\begin{equation}
  \nu(z) = \Frac{i}{\pi} \partial_{z^\ast} \lim_{\eta \to 0} \int
  \Frac{d \vect{r}}{L^d} \left. \Frac{\delta^2}{\delta
  \bar{j}_2^{\smc{b}} (\vect{r}) \delta j_1^{\smc{b}} (\vect{r})} Z
  [j] \right|_{j = 0} \; .
\label{eq:sourc}
\end{equation}

To be concrete, let us first consider the properties of the
Hamiltonian involving the imaginary scalar potential. To exploit the
analogy with the superconducting system, it is again convenient to
implement the gauge transformation $\psi \mapsto \Gamma \psi$
whereupon the generating functional assumes the form
\begin{equation}
  \mathcal{Z}[0]= \int D (\bar{\psi} , \psi) \exp \left[i
  \int d \vect{r} \; \bar{\psi} (\vect{r}) \left(i \eta -
  \hat{\mathcal{H}}_\Gamma\right) \psi (\vect{r}) \right] \; ,
\label{eq:gauge}
\end{equation}
where $\hat{\mathcal{H}}_\Gamma$ is defined above~\eqref{eq:gprime}. 
Now, as with the superconductor, the matrix
Hamiltonian~\eqref{eq:vdoub} satisfies the particle/hole or `charge
conjugation' ($\smc{cc}$) symmetry
\begin{equation*}
  -\sigma_2 \hat{\mathcal{H}}_\Gamma^\mathsf{T} \sigma_2 =
  \hat{\mathcal{H}}_\Gamma \; .
\end{equation*}
The latter is responsible for quantum interference effects which
modify the long-range or low-energy properties of the average Green
function. To accommodate these effects it is convenient to double the
field space
\begin{equation*}
\begin{split} 
  2 \bar{\psi} \left(i \eta- \hat{\mathcal{H}}_\Gamma\right) \psi 
  &= \bar{\psi} \left(i \eta - \hat{\mathcal{H}}_\Gamma\right) \psi +
  \psi^\mathsf{T} \left(i \eta-
  \hat{\mathcal{H}}_\Gamma^\mathsf{T}\right) \bar{\psi}^\mathsf{T} \\
  &= \bar{\psi} \left(i \eta - \hat{\mathcal{H}}_\Gamma\right) \psi
  + \psi^\mathsf{T} \left(i \eta + \sigma_2 \hat{\mathcal{H}}_\Gamma
  \sigma_2\right) \bar{\psi}^\mathsf{T} \\
  &= 2 \bar{\Psi} \left(i \eta\sigma^{\smc{cc}}_3
  -\hat{\mathcal{H}}_\Gamma\right) \Psi \; ,
\end{split}
\end{equation*}
where, defining the Pauli matrix $\sigma_3^{\smc{cc}}$ which operates
in the charge-conjugation ($\smc{cc}$) space,
\begin{align*}
  \Psi        &= \Frac{1}{\sqrt{2}} 
  \begin{pmatrix} 
  \psi \\ 
  \sigma_2
  {\bar{{\psi}}}^\mathsf{T}  
  \end{pmatrix}_{\smc{cc}} &
  \bar{\Psi}  &= \frac{1}{\sqrt{2}} 
  \begin{pmatrix} 
  \bar{\psi} & -\psi^\mathsf{T} \sigma_2
  \end{pmatrix}_{\smc{cc}} \; .
\end{align*}
As a result, the generating functional takes the form:
\begin{equation*}
  \mathcal{Z}[0] = \int D(\bar{\Psi},\Psi) \exp\left\{i\int d\vect{r}
  \; \bar{\Psi} (\vect{r}) \left[i \eta \sigma_3^{\smc{cc}} +\left(x
  - \hat{\zeta}_{\hat{\vect{p}}} - W\right) \sigma_3 -\left(y -
  V\right) \sigma_2\right] \Psi (\vect{r})\right\} \; . 
\end{equation*}
With this definition, the fields $\Psi$ and $\bar{\Psi}$ are not
independent but obey the symmetry relations
\begin{align}
  \Psi &= \sigma_2 \gamma \bar{\Psi}^{\mathsf{T}} & \bar{\Psi} &= -
  \Psi^{\mathsf{T}} \sigma_2 \gamma^{\mathsf{T}} \; ,
\label{eq:symme}
\end{align}
with $\gamma = E_{\smc{bb}} \otimes \sigma_1^{\smc{cc}} - i
E_{\smc{ff}} \otimes \sigma_2^{\smc{cc}}$, where $E_{\smc{bb}} = \diag
(1 , 0)_{\smc{bf}}$ and $E_{\smc{ff}} = \diag (0 , 1)_{\smc{bf}}$ are
the projectors on the boson-boson and fermion-fermion sector
respectively.
This completes the construction of the generating functional as a
functional field integral. To make further progress, we can draw on
the intuition afforded by existing studies of the superconducting
system. 

\subsubsection{Disorder Averaging}
\label{sec:disor}
As a first step towards the construction of an effective low-energy
theory it is necessary to subject the generating functional to an
ensemble average over the random impurity distribution. For this
purpose we will take both the real and imaginary components of the 
random scalar potential to be Gaussian $\delta$-correlated with zero mean, 
and correlation 
\begin{align}
  \langle W (\vect{r}_1) W (\vect{r}_2)\rangle_W &= \Frac{1}{2
  \pi\nu\tau} \delta^d (\vect{r}_1 - \vect{r}_2) & \langle V
  (\vect{r}_1) V (\vect{r}_2)\rangle_V &= \Frac{1}{2 \pi \nu \tau_n}
  \delta^d (\vect{r}_1 - \vect{r}_2) \; ,
\label{eq:diswv}
\end{align}
where $\nu\equiv(\Delta L^d)^{-1} \sim m (2 m x)^{(d-2)/2}$ represents
the unperturbed DoS of the Hermitian Hamiltonian 
$\hat{\zeta}_{\hat{\vect{p}}}$ at energy $x$, while $\tau$ and $\tau_{n}$ 
denote, respectively, the mean scattering time of the real and imaginary 
components of the potential. In the following, we will suppose that the 
imaginary component of the scattering potential is weak, i.e. 
$\tau_{n}\gg \tau$.

The ensemble averaging over the real random scalar potential
$W(\vect{r})$ induces an interaction of the fields which can be
decoupled by the introduction of an $8\times 8$ component 
Hubbard-Stratonovich field $Q$, 
\begin{displaymath}
  \langle \exp \left[-i \int d \vect{r} \; \bar{\Psi} W \sigma_3
  \Psi\right] \rangle_W = \int D Q \exp \left[\int d \vect{r}
  \left(\Frac{\pi \nu}{8 \tau} \str Q^2 - \Frac{1}{2 \tau} \bar{\Psi}
  Q \sigma_3 \Psi\right) \right] \; .
\end{displaymath}
Transforming the fields according to the symmetry~\eqref{eq:symme},
one finds that the supermatrix fields $Q(\vect{r})$ must be subjected
to the constraint:
\begin{equation}
  Q = \sigma_1 \gamma Q^{\mathsf{T}} \gamma^{\mathsf{T}} \sigma_1 \;
  . 
\label{eq:qsymm}
\end{equation}
Similarly, the interaction of the fields induced by the ensemble
average over the random imaginary scalar potential $V(\vect{r})$ can
be decoupled by an $8\times 8$ component Hubbard-Stratonovich field $P
(\vect{r})$, satisfying the symmetry constraint, $P= \gamma
P^{\mathsf{T}} \gamma^{\mathsf{T}}$:
\begin{displaymath}
  \langle \exp \left[i \int d \vect{r} \; \bar{\Psi} V \sigma_2
  \Psi\right] \rangle_V = \int D P \exp \left[\int d \vect{r}
  \left(\Frac{\pi \nu}{8 \tau_{n}} \str P^2 - \Frac{1}{2 \tau_n}
  \bar{\Psi} P \sigma_2 \Psi\right) \right] \; .
\end{displaymath}
Finally, an integration over the fields $\Psi$ obtains
\begin{gather*}
  \langle \mathcal{Z} [0] \rangle_{W,V} = \int DQ \int DP \exp
  \left\{\Frac{\pi \nu}{8} \int d\vect{r} \;
  \str\left(\Frac{Q^2}{\tau} + \Frac{P^2}{\tau_n}\right) - \Frac{1}{2}
  \int d\vect{r} \; \str \langle \vect{r} | \ln \hat{\mathcal{G}}^{-1} |
  \vect{r} \rangle \right\} \; ,
\intertext{where}
  \hat{\mathcal{G}}^{-1} = i \eta \sigma_3 \otimes \sigma_3^{\smc{cc}}
  + x - \hat{\zeta}_{\hat{\vect{p}}} - iy\sigma_1 + \Frac{i}{2 \tau} Q
  - \Frac{1}{2 \tau_n} P \sigma_1 
\end{gather*}
represents the supermatrix Green function. Further progress is
possible only within the a saddle-point approximation. Following 
Ref.~\cite{altland_simons}, in the quasi-classical approximation
$x \gg 1/\tau \gg [1/\tau_n , y]$, it is convenient to implement a
two-level procedure in which we first formulate an intermediate energy
scale action in which the terms in $y$ and $1/\tau_n$ are treated as
small symmetry breaking sources. Later, the influence of $V$ (through
$P$) on the complex DoS can be explored within a  stationary analysis
of this reduced action.

\subsubsection{Intermediate Energy Scale Action}
\label{sec:inter}
Neglecting the terms in $y$ and $P$, a variation of the action with 
respect to $Q$ obtains the saddle-point equation 
\begin{align*}
  Q(\vect{r}) &= \frac{i}{\pi \nu} \mathcal{G}_0 (\vect{r} ,
  \vect{r}) &
  \hat{{\mathcal{G}}}_0^{-1} &= i \eta
  \sigma_3^{\smc{cc}}\otimes\sigma_3 + x -
  \hat{\zeta}_{\hat{\vect{p}}} + \Frac{i}{2 \tau} Q \; .
\end{align*}
In the quasi-classical limit, $x \tau \gg 1$, taking $Q$ to be
homogeneous in space, this equation can be solved in the pole
approximation with $Q=\pm 1$. Taking into account the analytical
properties of the average Green function, we identify the solution
$Q_{\text{sp}} = \sigma_3 \otimes \sigma_3^{\smc{cc}}$. However, in
the limit $\eta \to 0$, the saddle-point solution is not unique, but
expands to span a degenerate manifold of solutions generated by the
homogeneous pseudo-unitary transformations $Q_{\text{sp}} =T\sigma_3
\otimes \sigma_3^{\smc{cc}} T^{-1}$ compatible with the symmetry of
$Q$, \eqref{eq:qsymm}.

Fluctuations around the saddle-point can be classified into longitudinal 
and transverse modes according to whether or not they violate the 
non-linear constraint $Q^2 (\vect{r})= \mathbb{I}$. In the quasi-classical
limit, the longitudinal fluctuations are rendered massive and do not 
contribute to the low-energy properties of the generating functional. 
On the other hand, transverse fluctuations of the fields become soft and 
must be accommodated. 
Taking into account soft fluctuations $Q (\vect{r}) = T(\vect{r}) \sigma_3 
\otimes \sigma_3^{\smc{cc}} T^{-1}(\vect{r})$, and restoring the 
perturbations $y$ and $P$, a gradient expansion of the action obtains the 
average generating functional
\begin{multline*}
  \langle \mathcal{Z} [0] \rangle_{W,V} = \int_{Q^2=\mathbb{I}} DQ
  \int DP \exp \left\{\Frac{\pi\nu}{8\tau_n} \int d \vect{r}\; \str
  P^2\right\} \\
  \exp\left\{\Frac{\pi \nu}{8} \int d\vect{r}\; \str \left[D (\nabla
  Q)^2 + 4 i \left( i \eta \sigma_3 \otimes \sigma_3^{\smc{cc}} - iy
  \sigma_1 - \Frac{P}{2\tau_n} \sigma_1 \right) Q \right]\right\}\; ,
\end{multline*}
where $D=v_F^2\tau/d$ is the diffusion constant. Here we have
neglected terms higher order in $y\tau$ and $\tau/\tau_n$. In the same
approximation $y\tau \ll 1$, the supermatrix Green function takes the 
form~\cite{efetov} 
\begin{equation*}
  \mathcal{G}_0 (\vect{r} , \vect{r}') = -i \pi \nu f_d (|\vect{r} -
  \vect{r}'|) Q\left(\Frac{\vect{r} + \vect{r}'}{2}\right) \; ,
\end{equation*}
where $f_d(r)= \Gamma(d/2) [2/k_F r]^{d/2 -1} J_{d/2 -1} (k_F
r) \exp [- r/2 \ell]$ is the Friedel function, and $k_F=\sqrt{2mx}$.

Finally, integrating over the supermatrices $P$, we obtain the average 
generating functional
\begin{gather}
\notag
  \langle \mathcal{Z} [0] \rangle_{W,V} = \int_{Q^2=\mathbb{I}} DQ
  \exp \left\{-S[Q]\right\} \; ,
\intertext{where}
  S[Q] = -\Frac{\pi \nu}{8} \int d \vect{r} \; \str \left[D (\nabla
  Q)^2 -4 \left(\eta \sigma_3 \otimes \sigma_3^{\smc{cc}} - y
  \sigma_1\right) Q +\Frac{1}{\tau_n} \left(\sigma_1 Q\right)^2
  \right] \; . 
\label{eq:deeft}
\end{gather}

This completes the construction of the intermediate energy scale
action for the non-Hermitian system. Returning to the analogy with the 
superconductor, we note that the form of the action~\eqref{eq:deeft} 
coincides with that of a disordered superconductor subject to a random
real order parameter with an average magnitude $y$ and variance set by
$1/\tau_n$. To investigate the influence of the imaginary potential on
the complex DoS, it is necessary to subject this action to a further
stationary analysis and obtain the low-energy action.

\subsubsection{Low-Energy Scale Action}
\label{sec:lowen}
Although the soft mode action~\eqref{eq:deeft} is stabilized by the
quasi-classical parameter $x\tau$, the terms in $y$ and $1/\tau_n$
render the majority of field fluctuations massive. To obtain the
low-energy saddle-point, let us again vary the action with respect to
$Q$ subject to the non-linear constraint, $Q^2=\mathbb{I}$:
\begin{equation*}
  D \nabla (Q \nabla Q) -[\eta \sigma_3 \otimes
  \sigma_3^{\smc{cc}} - y \sigma_1 , Q] + \Frac{1}{2
  \tau_n} [\sigma_1 Q \sigma_1 , Q] = 0 \; . 
\end{equation*}
Applying the \emph{Ansatz} $Q = \cos\hat{\theta} \sigma_3 \otimes 
\sigma_3^{\smc{cc}} +\sin \hat{\theta} \sigma_1$, where $\hat{\theta} = 
\diag (i\theta_{\smc{bb}} , \theta_{\smc{ff}})_{\smc{b,f}}$, the saddle-point
equation assumes the form 
\begin{equation}
  D \nabla^2 \hat{\theta} - 2 \left[ \eta \sin \hat{\theta} + y \cos
  \hat{\theta} \right] - \Frac{1}{\tau_n} \sin 2 \hat{\theta} =
  0 \; ,
\label{eq:ineqn}
\end{equation}
a result reminiscent of the Abrikosov-Gor'kov mean-field equation of
the disordered superconductor subject to a pair-breaking
perturbation~\cite{abrikosov_gorkov}. Taking the solution to be
symmetric and homogeneous in space, one obtains
\begin{equation}
  i\theta_{\smc{bb}}=\theta_{\smc{ff}}=\theta_{\smc{mf}} = 
  \begin{cases}
  - \arcsin (y \tau_n)           & |y| \tau_n < 1 \\[.3cm]
  - \Frac{\pi}{2} \sign (y)      & |y| \tau_n \ge 1 \; .
  \end{cases}
\label{eq:mfsol}
\end{equation}

Once again, as $\eta \to 0$, this solution is not unique but expands
to span an entire manifold parameterized by transformations $Q =
T Q_{\smc{mf}} T^{-1}$, where $[T , \sigma_1] =0$ and $T = \gamma
(T^{-1})^{\mathsf{T}} \gamma^{\mathsf{T}}$ --- these represent the
soft fluctuations corresponding to the time-reversal invariant
superconductor symmetry class CI. Substituted back into the action,
these fluctuations are described by the corresponding low-energy
action~\cite{izyumov_simons1}
\begin{equation*}
  S_{\text{eff}} [\bar{Q}] = - \Frac{\pi \nu}{8} \int d \vect{r} \; D
  \left(1 - y^2 \tau_n^2\right) \str (\nabla \bar{Q})^2 \; ,
\end{equation*}
where $\bar{Q} = T \sigma_3 \otimes \sigma_3^{\smc{cc}} T^{-1}$. 
This completes our construction of the field theory describing
spectral correlations in the disordered non-Hermitian system. In the
following section we will use these results to explore the support of
the complex DoS.

\begin{figure}
\begin{center}
\includegraphics[width=1\linewidth,angle=0]{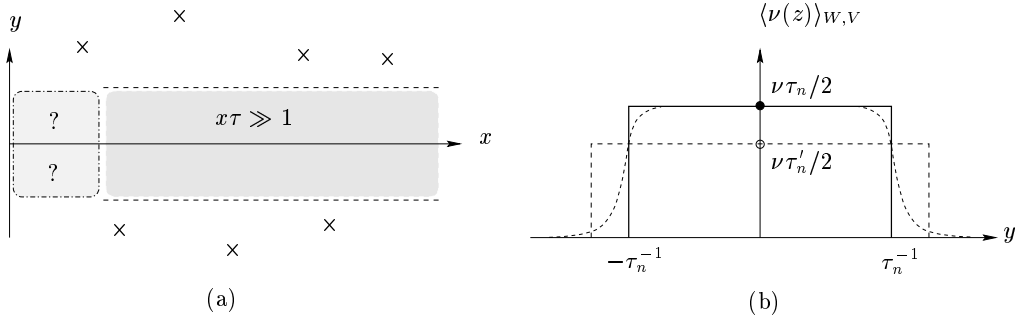}
\end{center}
\caption{\small
        Complex DoS in the quasi-classical limit $x \tau\gg 1$ for the 
        two-dimensional scalar potential problem. (a) In the
        mean-field approximation, the DoS is equal to $\nu \tau_n /2$
        inside the region $|y| \tau_n \le 1$ of the complex plane $z =
        x + i y$ and zero outside; (b) $\langle \nu (z) \rangle_{W,V}$
        versus $y$ for a fixed $x$ and for two values of the
        scattering time, $\tau_n > \tau_n^\prime$: while in the
        mean-field approximation the DoS shows sharp edges, rare
        realizations of the random potential give rise to tail states
        (\S~\ref{sec:state}).}
\label{fig:fidos}
\end{figure}

\subsection{Complex Density of States} 
\label{sec:state}
Making use of Eq.~\eqref{eq:sourc}, the complex DoS is obtained from 
the generating function as 
\begin{equation}
  \left\langle \nu (z , \vect{r}) \right\rangle_{W,V} =
  -\Frac{i}{4}\nu \partial_{z^\ast} \lim_{\eta \to 0} \langle \str 
  \left[\left(\mathbb{I} - \sigma_1\right) \otimes E_{\smc{cc}}^{11}
  \otimes \sigma_3^{\smc{bf}} Q(\vect{r})\right] \rangle_Q \; ,
\label{eq:nu_gen}
\end{equation}
where $\langle \dots  \rangle_Q = \int_{Q^2 =\mathbb{I}} DQ \dots 
e^{-S[Q]}$ where $S[Q]$ represents the intermediate energy scale
action~\eqref{eq:deeft}. In the homogeneous saddle-point
(i.e. mean-field) approximation~\eqref{eq:mfsol}, the corresponding
DoS takes the form
\begin{equation}
  \langle \nu(z) \rangle_{W,V} = i \nu \partial_{z^\ast}
  \lim_{\eta \to 0} \sin \theta_{\smc{mf}} =
  \begin{cases} 
  \Frac{\nu \tau_n}{2}  & \quad |y| \tau_n <
  1\\[.3cm] 
  0                           & \quad |y| \tau_n \ge 1\; .
  \end{cases}
\label{eq:mfdos}
\end{equation}
As expected, at the level of the mean-field, the field theory predicts 
a migration of the DoS off the real line and into the complex
plane. The density of complex eigenvalues is constant over the region
of support set by the effective scattering rate of the non-Hermitian
potential. Reassuringly, the expression for the DoS satisfies the sum
rule $\int dy \langle \nu(z) \rangle_{W,V} = \nu$. In the
two-dimensional case, the result is illustrated qualitatively in
Fig.~\ref{fig:fidos}. As the scattering time $\tau_n$ does not depend
on the real part of the energy $x$, the width of the mean-field
spectrum boundary is constant.

To complete the analysis, it is necessary to take into account the 
fluctuations of the fields around the saddle-point. As discussed above,
these fluctuations divide into a set which are massive (on the scale 
of $y$) and a massless set. Interestingly, one finds that the latter 
commute with the source and do not contribute to the DoS! 

Indeed, the general profile of the support is born out by numerical 
simulation. Figure~\ref{fig:specs} shows the amalgamation of data for
the two dimensional random system generated from $100$ realizations
of the complex random scalar potential Hamiltonian. However, instead
of a sharp cut-off in the support, the data clearly indicate a soft
edge with eigenvalues occupying the region outside the mean-field
prediction~\eqref{eq:mfdos}. How can this observation be reconciled
with the results of the field theory? Surprisingly, taking into
account massive fluctuations of the action perturbatively, the
mean-field prediction remains intact. In fact, the tail states that
decorate the edge of the support are generated by optimal fluctuations
of the \emph{real} random impurity potential and are reflected in
\emph{non-perturbative} instanton configurations of the action.

\begin{figure}
\begin{center}
\includegraphics[width=0.7\linewidth,angle=0]{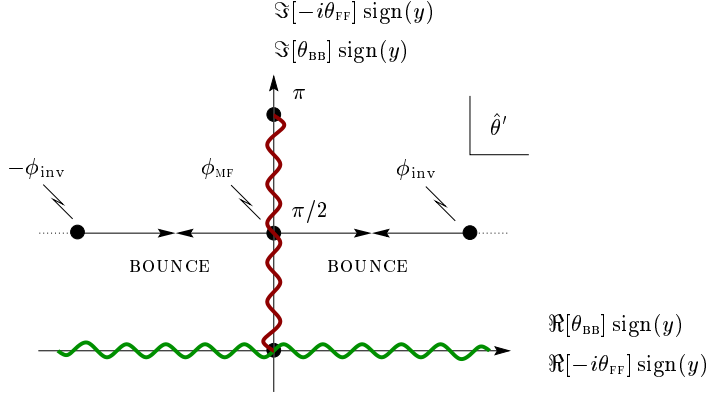}
\end{center}
\caption{\small
        Integration contours for boson-boson and fermion-fermion
        fields in the complex plane $\hat{\theta'} = - i \hat{\theta}
        \sign (y)$. The two degenerate bounce solutions are
        qualitatively shown, together with the mean-field starting
        point $\phi_{\smc{mf}}$ and the inversion point
        $\phi_{\text{inv}}$.
        }
\label{fig:conto}
\end{figure}

\subsection{Tail States and Instantons} 
To identify corrections to the mean-field DoS~\eqref{eq:mfdos} we can
draw on the intuition afforded by recent studies of sub-gap state
formation due to optimal fluctuations in the superconducting
system~\cite{lamacraft_simons1}.  
While the homogeneous solution of the mean-field equation gives rise to a 
hard edge of the DoS support, we will see that inhomogeneous symmetry broken 
field configurations reflect the influence of rare realizations or 
`optimal fluctuations' of the random scalar potentials which soften 
the edge (Fig.~\ref{fig:fidos}). Here, for simplicity, let us focus on 
the quasi one-dimensional case, generalizing the discussion to the 
$d$-dimensional system in section~\ref{sec:highd}.

\subsubsection{Quasi One-Dimensional Geometry}
\label{sec:quasi}
To explore the influence of inhomogeneous field configurations of the
action, let us revisit the mean-field equation of motion~\eqref{eq:ineqn}. 
Operationally, it is convenient to deal not with the saddle-point 
equation~\eqref{eq:ineqn} itself, but rather its first integral, 
$D (\nabla \hat{\theta})^2 - y V(\hat{\theta}) =\text{const}$, where
\begin{equation*}
V (\hat{\theta}) = - 4 \left[ \Frac{\eta}{y} \cos \hat{\theta} - \sin
\hat{\theta}\right] - \Frac{1}{y \tau_n} \cos 2\hat{\theta} \; ,
\end{equation*}
denotes the effective complex `potential'. 

\begin{figure}
\begin{center}
\includegraphics[width=0.6\linewidth,angle=0]{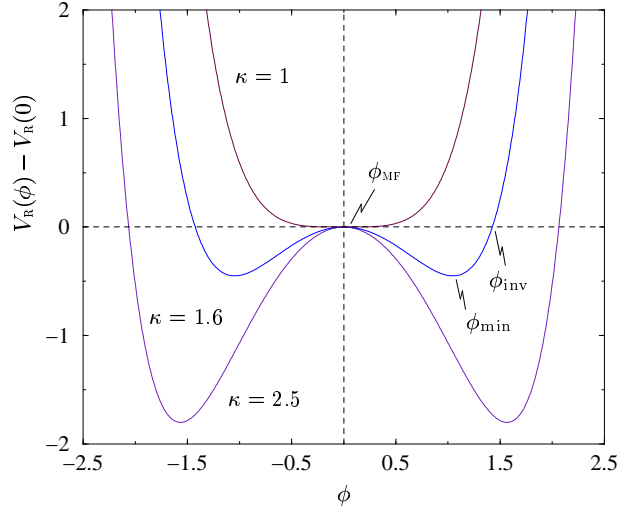}
\end{center}
\caption{\small
        Rescaled potential $V_{\smc{r}} (\phi) - V_{\smc{r}} (0)$
        versus the variable 
        $\phi$ for  $\kappa = 2.5$, $\kappa = 1.6$ and $\kappa =
        1$.}
\label{fig:poten}
\end{figure}

To identify the inhomogeneous instanton solution let us first note
that, outside the region of support, $|y| \tau_n \ge 1$, the 
homogeneous mean-field solution takes the form $\theta_{\smc{mf}} =
- \pi/2 \sign (y)$.
Taking into account the condition that the solution should coincide
with $\theta_{\smc{mf}}$ at infinity, one can identify a `bounce'  
solution parameterized by $\theta=[-\pi/2+i\phi]\sign (y)$, with
$\phi$ real, involving the \emph{real} potential
$V_{\smc{r}}(\phi) \equiv V([-\pi/2+i\phi] \sign (y))$ with endpoint
$\phi_{\text{inv}}$ such that $V_{\smc{r}}(\phi_{\text{inv}}) =
V_{\smc{r}} (\phi_{\smc{mf}})$.

Now integration over the angles $\hat{\theta}$ is constrained to
certain contours~\cite{efetov}. Is the bounce solution accessible to
both? As usual, the contour of integration over the boson-boson field
$\theta_{\smc{bb}}$ includes the entire real axis, while for the
fermion-fermion field $i\theta_{\smc{ff}}$ runs along the imaginary
axis from $0$ to $i\pi$. With a smooth deformation of the integration
contours, the mean-field saddle-point is accessible to both the angles
$\hat{\theta}$~\cite{altland_simons,bundschuh}. By contrast, the
particular bounce solution 
\emph{and} the mean-field solution can be reached simultaneously by a
smooth deformation of the integration contour \emph{only} for the
boson-boson field $\theta_{\smc{bb}}$ (see Fig.~\ref{fig:conto}). The
particular bounce solution breaks supersymmetry.

Applying the parameterization for the `bounce'
\begin{align*}
  i \theta_{\smc{bb}} (r) &= \left[ - \Frac{\pi}{2} + i \phi
  (r) \right] \sign (y) & \theta_{\smc{ff}} = \theta_{\smc{mf}}
  \; ,
\end{align*}
the first integral takes the form
\begin{gather}
\begin{align}
  (\nabla_{r/\xi} \phi)^2 + V_{\smc{r}} (\phi) &= V_{\smc{r}}
  (\phi_{\smc{mf}}) &
  V_{\smc{r}} (\phi) &= -4 \cosh \phi + \kappa^{-1} \cosh 2 \phi \; ,
\label{eq:first}
\end{align}
\intertext{where $\xi=(D/|y|)^{1/2}$ and}
\kappa = \tau_n |y|\; .
\label{eq:newco}
\end{gather}
Here $\phi_{\smc{mf}} = 0$ for $|y| \tau_n \ge 1$, while
$\phi_{\smc{mf}}$ is purely imaginary for $|y| \tau_n <
1$. Typical shapes of the potential $V_{\smc{r}} (\phi)$ for different values
of $\kappa$ are shown in Fig.~\ref{fig:poten}. A bounce solution
exists only for values of $\kappa$ bigger than one, while for $\kappa
< 1$ the unique solution is the homogeneous one, $\phi =
\phi_{\smc{mf}}$.

\begin{figure}
\begin{center}
\includegraphics[width=0.6\linewidth,angle=0]{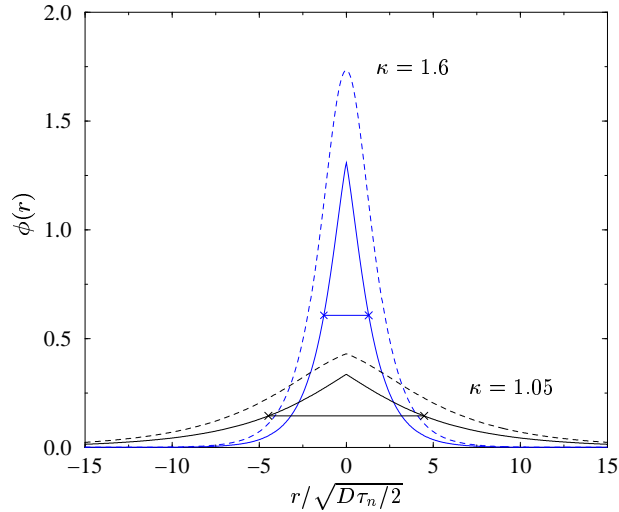}
\end{center}
\caption{\small
        Exact (solid line) and approximate (dashed line) bounce
        solution $\phi (r)$ versus $r/\sqrt{D \tau_n /2}$ in the quasi
        one-dimensional geometry for two different values of the
        inverse of the imaginary `impurity' concentration, $\kappa =
        1.6$ and $\kappa = 1.05$. The interval $[-r_0 (\kappa) , r_0
        (\kappa)]$ is explicitly plotted for the exact solutions.
        }
\label{fig:phifu}
\end{figure}

\begin{figure}
\begin{center}
\includegraphics[width=0.6\linewidth,angle=0]{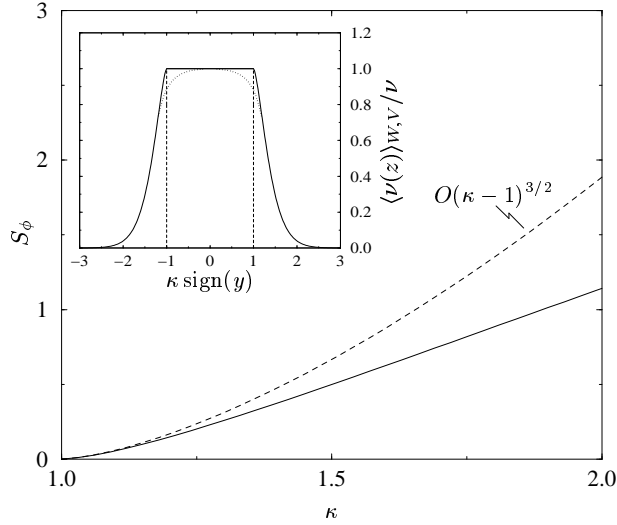}
\end{center}
\caption{\small
        Exact action $S_\phi$ (solid line) versus $\kappa$ in the
        quasi one-dimensional geometry together with the first term of
        the expansion in $(\kappa - 1)$ (dashed line). The action
        vanishes as $\kappa \to 1^+$. In the inset the rescaled DoS
        $\langle \nu (z) \rangle_{W,V} / \nu \sim \exp [-
        S_{\text{inst}}]$ versus $y \tau_n$, for
        $\pi \nu \sqrt{D/\tau_n} = 2$, is
        plotted. For values of $\kappa \sign (y)$ near the critical
        values $\pm 1$, the saddle-point analysis becomes unreliable. 
        Here, fluctuations lead to a smooth interpolation between the 
        DoS in the tail region, and that coming from the bulk mean-field. 
        }
\label{fig:expsi}
\end{figure}

For the quasi one-dimensional system it is possible to derive analytic 
expressions for both the real instanton action $S_{\text{inst}} =
2 \pi \nu |y| \xi S_\phi (\kappa)
$
\begin{equation}
  S_\phi (\kappa) = \int_{0}^{\phi_{\text{inv}}} d \phi \sqrt{V_{\smc{r}} (0)
  - V_{\smc{r}} (\phi)} \; , 
\label{eq:inact}
\end{equation}
where $\phi_{\text{inv}} = |\arccosh ( 2 \kappa -1 )|$ is the endpoint
of the motion, and the bounce solution $\phi (r)$. In particular,
for $S_\phi (\kappa)$ we have (Fig.~\ref{fig:expsi})
\begin{equation}
  S_{\phi} (\kappa) = \sqrt{2 \kappa} \left\{ - 2 \Frac{\sqrt{\kappa - 
  1}}{\kappa} + 2 \arctan \sqrt{\kappa - 1} \right\} \; .
\label{eq:actio}
\end{equation}
At the same time, imposing the boundary conditions $\phi (r \to \pm
\infty) = 0$, from the first integral~\eqref{eq:first} one obtains the
bounce solution
\begin{gather}
\notag
  \cosh \phi(r) = \Frac{\kappa^2 + 2 e^{-2|r|/r_0(\kappa)}\sqrt{\kappa
  - 1} (3 \kappa - 2) + e^{-4|r|/r_0(\kappa)}(\kappa - 1)}{\kappa^2 +
  2 e^{-2|r|/r_0(\kappa)} \sqrt{\kappa - 1} (2 - \kappa) +
e^{-4|r|/r_0(\kappa)}(\kappa - 1)} \; , \\
\intertext{where}
  r_0(\kappa)= \left[ \Frac{D \tau_n}{2 (\kappa - 1)}
  \right]^{1/2} 
\label{eq:dropl}
\end{gather}
defines the extent of the `droplet' (see Fig.~\ref{fig:phifu}). In 
particular, we note that, one approaching the band edge ($\kappa \to
1^+$), $r_0(\kappa) \to \infty$.

Now close to the band edge ($\kappa\to 1^+$), the maximum of the potential
$V_{\smc{r}}(\phi)$ converges on the two minima allowing for the development
of a perturbative expansion in $\phi \simeq\phi_{\smc{mf}} = 0$. Setting
$\beta = (4  - \kappa)/6 \kappa \simeq 1 / 2$,
\begin{equation}
  V_{\smc{r}} (\phi) \Simiq_{\phi \simeq 0} V_{\smc{r}} (0) - 2 \Frac{\kappa -
  1}{\kappa} \phi^2 + \beta \phi^4 + O(\phi^6) \; .
\label{eq:devel}
\end{equation}
In this approximation the action and the bounce solution are given
respectively by:
\begin{align*}
  S_{\phi} (\kappa) &\Simiq_{\kappa \gtrsim 1} \Frac{4
  \sqrt{2}}{3} (\kappa - 1)^{3/2} & \phi(r) &\Simiq_{\kappa 
  \gtrsim 1} 4 \sqrt{2} \sqrt{\Frac{\kappa - 1}{\kappa}} \Frac{ 
  e^{-|r|/r_0(\kappa)}}{e^{-2|r|/r_0(\kappa)} +
  4 \beta} \; .
\end{align*}

This completes the estimate of the contribution to the action from a 
single bounce configuration. However, to complete the analysis it
is necessary to explore the influence of fluctuations around the instanton
solution. Here we sketch the important aspects of the analysis referring 
for details to the parallel discussion by Ref.~\cite{lamacraft_simons1} 
in the context of the disordered superconductor. 
Generally, field fluctuations around the instanton solution can be 
separated into `radial' and `angular' contributions. The former 
involve fluctuations of the diagonal elements $\hat{\theta}$, while the 
latter describe rotations including those Grassmann transformations which 
mix the $\smc{bf}$ sector. Both classes of fluctuations play an important
role. 

Dealing first with the angular fluctuations, supersymmetry breaking of the 
bounce is accompanied by the appearance of a Grassmann zero mode separated 
by an energy gap from higher excitations. This Goldstone mode restores the 
global supersymmetry of the theory. Crucially, this mode ensures that the 
saddle-point respects the normalization condition $\langle
\mathcal{Z}[0] \rangle_{W,V} =1$. Associated with radial fluctuations
around the bounce, there exists a zero mode due to translational
invariance of the solution, and a negative energy mode
(c.f. Ref.~\cite{coleman}). 

Combining these contributions, one obtains the following expression for 
the local complex DoS in the tail region,
\begin{equation}
  \langle \nu (z , r) \rangle_{W,V} \Simi_{\kappa > 1} i\nu
  \partial_{z^*}
  \left[ \cosh \phi (r) - \cosh \phi_{\smc{mf}} \right] |\chi_0
  (r)|^2 e^{- S_{\text{inst}}} \; ,
\label{eq:fldos}
\end{equation}
where $\chi_0 (r)$ represents the eigenfunction for the Grassmann zero
mode and $S_{\text{inst}}$ denotes the instanton action~\eqref{eq:actio}.
Thus, to exponential accuracy, the complex local DoS in the tail region
becomes non-zero only in the vicinity of the bounce. Mechanisms of 
quantum interference due to optimal potential fluctuations in the 
non-Hermitian system conspire to localize states in the tail regions
on a length scale $r_0(\kappa)$ greatly in excess of the wavelength 
$\lambda=1/k_F$. 

\subsubsection{Generalization to Dimensions $1<d<4$}
\label{sec:highd}
Close to the band edge ($\kappa \gtrsim 1$), a generalization of the
quasi one-dimensional results to higher dimensions can be developed by
dimensional analysis. From the expansion~\eqref{eq:devel} for the
potential $V_{\smc{r}}(\phi)$, the bounce configuration is shown to
have the scaling form
\begin{displaymath}
  \phi(\vect{r}) = \Frac{1}{\sqrt{\beta}} \left[\Frac{\xi}{r_0
  (\kappa)}\right] f(\vect{r}/ r_0 (\kappa)) \; ,
\end{displaymath}
which, in turn, implies the instanton action~\eqref{eq:inact}:
\begin{equation}
  S_{\text{inst}} \Simiq_{\kappa\gtrsim 1} 4 \pi a_d
  (\Delta\tau_n)^{(d - 2)/2} g^{d/2} \left[2 (\tau_n |y| -
  1)\right]^{(4 - d)/2} \; . 
\label{eq:ddima}
\end{equation}
Here $g=\nu D L^{d-2}$ and $\Delta=(\nu L^d)^{-1}$ represent
respectively the dimensionless conductance and average level spacing
of the Hermitian system (i.e. with $V=0$), and $a_d$ is a numerical
constant ($a_0 = 1/4$ and $a_1 = 1/3$)\footnote{\ \label{note2}
  Specifically $a_d = \int d \vect{u} [(\nabla_\vect{u} f)^2 + f^2
(\vect{u}) - f^4(\vect{u})]$.}. Later, in discussing the universality
of the results, we will return to consider the zero-dimensional
situation.

\begin{figure}
\begin{center}
\includegraphics[width=0.6\linewidth,angle=0]{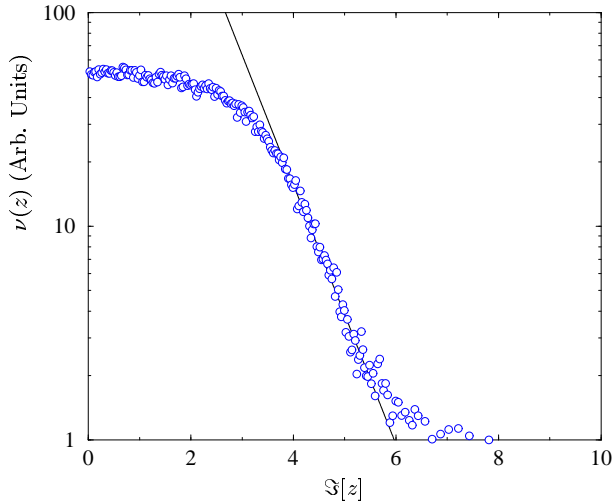}
\end{center}
\caption{\small
        Plot of the (logarithm of the) complex DoS versus $\Im [z] =
        y$ for a fixed value of $\Re [z] = x$ taken from the numerical
        simulation of Fig.~\ref{fig:specs}. The exponential fit of the
        data is shown as solid curve.}
\label{fig:dos_14}
\end{figure}

Taken together with Eq.~\eqref{eq:fldos}, this result shows that the 
complex DoS in the tail region varies exponentially with the
separation from the mean-field edge. The power law dependence of the
exponent depends on dimensionality, presenting a linear dependence on
$(\tau_n |y|-1)$ in the two-dimensional system. In this case,
moreover, the only dependence in~\eqref{eq:ddima} on the real part
of the energy $x$ occurs through the dimensionless conductance,
$g = \nu D \sim |x|$, so that the width of the exponential tail scales
as $1/|x|$, as qualitatively appears in Fig.~\ref{fig:specs}. A
comparison with the numerical simulation is shown in
Fig.~\ref{fig:dos_14}. The results show a good quantitative agreement
with the theory for the tail of the DoS. Note that close to (but
within) the edge of the support of the spectrum, the saddle-point
approximation used in estimating the mean-field DoS becomes
unreliable. As shown by the zero-dimensional analysis below, an honest
treatment of fluctuations in this region provides a smooth
interpolation between the bulk and tail states.

\section{Random Imaginary Vector Potential}
\label{sec:vecto}
To conclude our discussion of tail states in the non-Hermitian system,
we turn now to an analysis of the random imaginary vector potential 
Hamiltonian~\eqref{eq:hamiv}. Once again, our approach will be to 
formulate an effective field theory from which the spectral properties 
can be determined. Although, as shown in section~\ref{sec:backg}, the 
random vector potential Hamiltonian belongs to a different fundamental 
symmetry from the random scalar potential Hamiltonian, we will show below 
that, with some qualification, the DoS in the vicinity of the support 
exhibits the same universal scaling form. 

To be concrete, let us consider a $d$-dimensional 
Hamiltonian~\eqref{eq:hamiv} involving an imaginary vector potential
field $\vect{h}(\vect{r})$ together with a real random scalar
potential $W(\vect{r})$. Once again, let us assume that the scalar
potential $W(\vect{r})$ is drawn from a Gaussian $\delta$-correlated
impurity distribution~\eqref{eq:diswv}. Similarly, we will be
primarily concerned with a system exhibiting a random vector potential
drawn from a Gaussian distribution with zero mean $\langle
\vect{h}(\vect{r})\rangle_{h} = 0$ and correlation
\begin{equation}
  \langle h_i(\vect{q}) h_j(\vect{q}')\rangle_h = (2\pi)^d
  \delta(\vect{q} + \vect{q}') f (|\vect{q}|)\left[\gamma_1
  \left(\delta_{ij} - \Frac{q_i q_j}{\vect{q}^2}\right) + \gamma_2
  \Frac{q_i q_j}{\vect{q}^2}\right] \; .
\label{eq:corre}
\end{equation}
Here, for reasons that will become clear below, we have introduced an 
`envelope' $f(|\vect{q}|)$ which limits correlations of the fields at 
both microscopic and macroscopic length scales. Setting 
\begin{equation}
  \vect{h} (\vect{r}) = \nabla \varphi(\vect{r}) + \nabla \wedge
  \vectgr\chi (\vect{r}) \; ,
\label{eq:inair}
\end{equation}
$\gamma_1$ is identified as the strength of fluctuations of the 
incompressible component of the vector field, $\vectgr\chi(\vect{r})$, 
while $\gamma_2$ controls the irrotational part, $\varphi (\vect{r})$.
In the following, it will also be useful to contrast the behaviour of the 
random vector potential system with that of a \emph{constant} vector field 
$\vect{h} (\vect{r}) = \text{const} =\vect{h}_0$ --- the Hatano-Nelson
system~\cite{hatano_nelson}. 

\subsection{Background}

Before turning to the formal analysis of the statistical field theory, 
which begin with some general considerations which identify certain
idiosyncrasies of the constant vector potential system and how they 
impact upon the existence of tail states in the random vector potential 
system. Previous investigations of the former have revealed unusual 
localization properties of the random Hamiltonian which contrast with 
those of its Hermitian (i.e. $\vect{h}_0=0$) counterpart. Specifically, 
it was shown in Ref.~\cite{hatano_nelson} that, for the strictly 
one-dimensional system subject to {\em periodic boundary conditions} at 
infinity, when the field, $|\vect{h}_0|$, exceeds a critical value $h_c$ 
corresponding to the (energy dependent) localization length 
$\xi_{\text{loc}}=1/h_c$ of the Hermitian system, there is a transition
to a delocalized phase. Since then, this result has been generalized
to the quasi one-dimensional system by Kolesnikov and 
Efetov~\cite{kolesnikov_efetov} using techniques which parallel those
discussed here.

Now, following Ref.~\cite{hatano_nelson}, a simple argument can be
established to describe qualitatively the origin of the
transition. For a constant imaginary vector potential, $\vect{h}_0$, a
similarity transformation of the left and right eigenfunctions
$\phi^L_i(\vect{r}) \mapsto \phi_i^L(\vect{r}) \exp(-\vect{h}_0 \cdot
\vect{r})$ and $\phi^R_i (\vect{r}) \mapsto \exp(\vect{h}_0 \cdot
\vect{r}) \phi_i^R(\vect{r})$ removes from the Hamiltonian its
dependence on $\vect{h}_0$. Being now Hermitian, the eigenvalues of
the Hamiltonian must therefore be real and specified by those of the
unperturbed random system. However, if periodic boundary conditions
are imposed, the similarity transformation must be applied with
caution: if the eigenfunctions of the Hermitian system are localized
on a length scale $\xi_{\text{loc}}<|\vect{h}_0|$ the similarity
transformation is incompatible with the periodic boundary conditions
and can not be imposed. Here there is a transition from a localized
phase where the eigenvalues are real to a delocalized phase where the
eigenvalues migrate into the complex plane.

Now what does this phenomenology tell us about the existence of tail 
states? Using the same arguments, it is easy to deduce that optimal 
fluctuations of the impurity potential can {\em not} induce localized 
tail states in the constant vector potential system: suppose that a 
tail state appeared at some complex energy outside the band of bulk 
states. If the state is localized, it must be insensitive to the 
boundary conditions at infinity and, therefore, its dependence on the 
constant field can be removed by a similarity transformation. As such, 
it must therefore be contained within the Hermitian theory, and it's 
eigenvalue must be real. This contradiction strictly eliminates tail 
states from the infinite constant imaginary vector potential system.
(Of course, the Hermitian system can and will exhibit low-energy 
Lifshitz band tail states.)

With this background, let us now turn to consider the random vector 
potential system. Once again, for a random field $\vect{h} (\vect{r})$ 
drawn from the Gaussian distribution~\eqref{eq:corre}, it is 
straightforward to confirm that the similarity transformation
$\phi^L_i(\vect{r}) \mapsto \phi_i^L(\vect{r}) \exp[-\varphi(\vect{r})]$ 
and $\phi^R_i (\vect{r}) \mapsto \exp[\varphi(\vect{r})] 
\phi_i^R(\vect{r})$ removes from the Hamiltonian the dependence on the 
incompressible component, $\varphi(\vect{r})$. Therefore, when subject 
to any {\em purely} irrotational vector field $\vect{h}(\vect{r})=
\nabla\varphi(\vect{r})$ (with zero average), the eigenvalues remain real 
and coincide with those of the Hermitian system. Conversely, the 
incompressible field distribution $\vect{h} (\vect{r}) =\nabla \wedge 
\vectgr\chi (\vect{r})$ can not be removed by similarity transformation.
Such field configurations necessarily generate states with complex 
eigenvalues.

\begin{figure}
\begin{center}
\includegraphics[width=0.6\linewidth,angle=0]{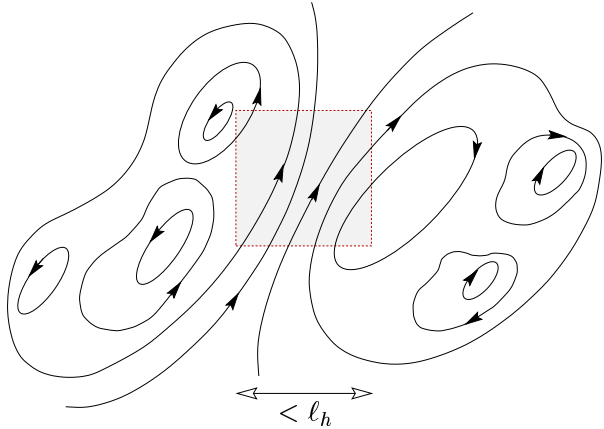}
\end{center}
\caption{\small
        A typical field configuration of the random vector field. 
        Over the interval $\ell_h$,
        the flux lines of $\vect{h}$ are approximatively constant,
        preventing the existence of tail state.
        }
\label{fig:qflux}
\end{figure}

Now, the random distribution~\eqref{eq:corre} with $f(|\vect{q}|)=1$ 
imposes long-ranged, i.e. power-law, correlations in the incompressible 
field configuration. As well as admitting superdiffusion 
processes~\cite{derrida,kravtsov,fisher} which dominate the relaxation 
to equilibrium in the classical system, these long-range correlations 
impact upon the probability of finding tail states in the non-Hermitian
system: any random configuration of the incompressible field will involve 
field lines correlated over arbitrarily long distances. These field
lines advect particles in the classical system and are responsible for
the characteristic superdiffusion properties of the system. Now in such a 
background, one can identify regions where the field lines are oriented
over long distances, say $\ell_h$ (see Fig.~\ref{fig:qflux}). Within this
region, the nucleation of tail states with a localization length $r_0<
\ell_h$ is prohibited by the same mechanism described above for the 
Hatano-Nelson system. Clearly, the phenomenology of tail state formation
in the superdiffusive system is subtle being sensitive to the arrangement
of both the scalar {\em and} vector field.

Therefore, since we are interested here in exposing the general principles 
behind tail state formation in the non-Hermitian system, to keep our 
discussion simple, we will limit our considerations to distributions 
where the correlations of the field lines are limited to a length scale 
$\ell_h$ much smaller than the localization length of the incipient 
tail states, $r_0$, i.e. introducing a microscopic cut-off $a$, we set 
\begin{displaymath}
  f(|\vect{q}|) = \begin{cases} 
  1 \quad & 1/\ell_h < |\vect{q}| < 1/a \\
  0 \quad & \text{otherwise} \; ,
  \end{cases}
\end{displaymath}
where $\ell_h$ represents the maximum length scale of field correlations.
With these considerations in mind, let us now turn to the development 
of the field theory of the non-Hermitian system. 

\subsection{Field Theory}

Following the method outlined for the imaginary scalar potential system, 
our starting point is the gauge transformed `Hermitized' matrix
Hamiltonian~\eqref{eq:vdoub}. Now, under the `charge-conjugation' 
operation, the Hamiltonian transforms as 
\begin{equation*}
  \sigma_2 \left[\hat{\mathcal{H}}_\Gamma (\vect{h})\right]^\mathsf{T}
  \sigma_2 = - \hat{\mathcal{H}}_\Gamma (-\vect{h}) \; .
\end{equation*}
The statistical properties of the non-Hermitian system can again be 
obtained from the generating function \eqref{eq:gauge} by forming an 
average over the real random potential $W (\vect{r})$ and introducing  
a slow field decoupling. In doing so, one obtains the generating
functional
\begin{gather}
\label{eq:trlog}
  \langle \mathcal{Z} [0] \rangle_W = \int DQ \exp \left[ \Frac{\pi
  \nu}{8 \tau} \int d\vect{r} \; \str Q^2 - \Frac{1}{2} \int d
  \vect{r} \; \str \langle \vect{r} | \ln \hat{\mathcal{G}}^{-1} |
  \vect{r} \rangle \right] \; ,
\intertext{where}
  \hat{\mathcal{G}}^{-1} = i \eta \sigma_3 \otimes \sigma_3^{\smc{cc}}
  + x - \Frac{1}{2m} \left[\hat{\vect{p}} - i \vect{h} (\vect{r})
  \sigma_1 \otimes \sigma_3^{\smc{cc}}\right]^2 - i y \sigma_1 +
  \Frac{i}{2 \tau} Q(\vect{r})
\notag
\end{gather}
denotes the supermatrix Green function. As with the scalar potential, 
further progress is possible only within a saddle-point approximation. 
However, following our discussion above, it is now necessary to exercise
some caution.

In the Hermitian system with a real vector potential $\vect{A}=i\vect{h}$, 
the conventional approach~\cite{efetov} involves subjecting the action to 
a saddle-point approximation in the absence of the field. The saddle-point 
manifold $Q^2=\mathbb{I}$ is unperturbed by the real vector potential 
and its effect on the low-energy properties of the system can be 
accommodated at the level of the gradient expansion. However, as 
emphasized in Ref.~\cite{kolesnikov_efetov}, in the presence of a 
\emph{constant} imaginary vector potential, $\vect{h}(\vect{r})=\vect{h}_0$, 
depending on the magnitude of $\vect{h}_0$, it is possible to identify 
separate saddle-points of the total effective action~\eqref{eq:trlog}. 
Inside the localized phase (i.e. $|\vect{h}_0| < h_c$), the dependence 
on $\vect{h}_0$ can (and must) be removed by a rotation of $Q$ --- the 
counterpart of the similarity transformation. The resulting theory 
reflects that of the Hermitian model and the eigenvalues condense onto 
the real line. Conversely, in the delocalized phase (i.e. $|\vect{h}_0| 
> h_c$), the similarity transformation on $Q$ is inadmissible. Here the 
dependence of the action on $\vect{h}_0$ must be developed explicitly, 
reflecting the fact that the eigenvalues acquire an imaginary component.
Similarly, for a random vector potential, it is necessary to remove the 
irrotational component of the field $\vect{h}$ (i.e. 
$\nabla\varphi(\vect{r})$) explicitly at the level of the saddle-point
by subjecting $Q$ to a similarity transformation
\begin{displaymath}
  Q (\vect{r}) \mapsto e^{ -\varphi (\vect{r}) \sigma_1
  \otimes \sigma_3^{\smc{cc}} }  Q(\vect{r}) e^{\varphi
  (\vect{r}) \sigma_1 \otimes \sigma_3^{\smc{cc}} } \; .
\end{displaymath}
---As expected, the same transformation leaves the source term for the 
DoS~\eqref{eq:nu_gen} unperturbed.

\begin{figure}
\begin{center}
\includegraphics[width=1\linewidth,angle=0]{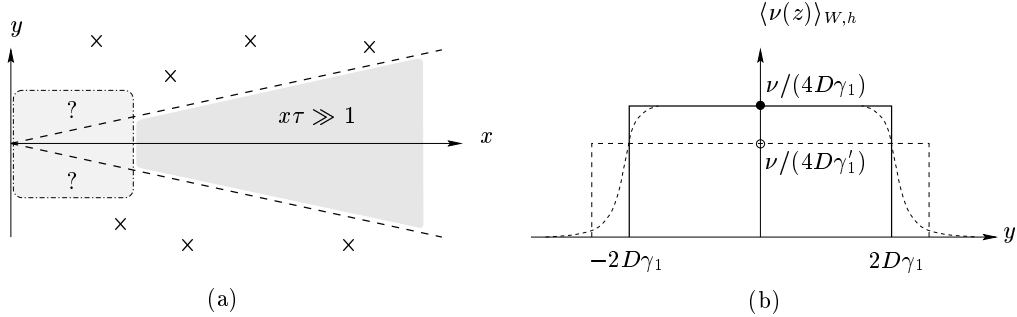}
\end{center}
\caption{\small
        Complex DoS for the imaginary vector potential problem in the
        quasi-classical limit $x \tau \gg 1$ for the two-dimensional
        system. (a) In the mean-field approximation the DoS is equal
        to $\nu /(4 D \gamma_1)$  inside the region $|y| \le 2 D
        \gamma_1 = 2 \tau \gamma_1 |x| /m$ of the complex plane $z = x
        + i y$ and zero otherwise; (b) $\langle \nu (z) 
        \rangle_{W,h}$ versus $y$ for two values of $\gamma_1 <
        \gamma_1^\prime$. The tails of the DoS for $|y| >2 \tau
        \gamma_1 |x| /m$ are shown schematically.}
\label{fig:vedos}
\end{figure}

Without the irrotational component of the field, the saddle-point of
the total action can be analyzed in the usual way, following a two-stage 
procedure. As before, taking into account the quasi-classical parameter 
$x\tau\gg 1$, one finds that the low-energy fluctuations are confined 
to the manifold $Q^2(\vect{r})=\mathbb{I}$. Then, subjecting the action 
to a gradient expansion keeping $y\tau\ll 1$ and $\sqrt{\gamma_1} \ll 
\ell^{-1}$, one obtains the non-linear $\sigma$-model 
action~\cite{efetov1,kolesnikov_efetov,efetov}
\begin{equation}
   S [Q] = -\Frac{\pi \nu}{8} \int d \vect{r} \; \str \left[D
  (\widetilde{\nabla} Q)^2 - 4 \left(\eta \sigma_3 \otimes
  \sigma_3^{\smc{cc}} - y \sigma_1\right) Q\right] \; ,
\label{eq:vnlsm}
\end{equation}
where $\widetilde{\nabla} = \nabla - \vect{h} [\ ,\sigma_1 \otimes
\sigma_3^{\smc{cc}}]$ represents the covariant derivative, and 
$\vect{h}\equiv \nabla \wedge \vectgr\chi$ reflects the incompressible
component of the field. 

Now, although the structure of the action compares with that of the 
imaginary scalar potential field, it exhibits important differences 
which substantially influences the behaviour of the system: specifically, 
in addition to the `diamagnetic' contribution, i.e. the term proportional 
to $\vect{h}^2$, the action presents a paramagnetic contribution, linear 
in $\vect{h}$, which couples to the gradient of $Q$. Although this term 
does not affect the homogeneous saddle-point solution, it is responsible 
for the destabilisation of the tail states in the system with long-range 
correlations. To see this explicitly, let us return to the consideration
of the constant vector potential system. Previously, in the qualitative 
discussion above, it was argued that such a system does not accommodate 
tail states. How is this phenomenon reflected in the field theory? 

In fact, for $\vect{h}(\vect{r}) = \vect{h}_0 > h_c$, the action is
still given by the non-linear
$\sigma$-model~\eqref{eq:vnlsm}. However, in view of the paramagnetic
term, to explore the saddle-point structure, it is
necessary to adopt a more general parameterization, $Q =
\sigma_1 \cos \hat{\theta} + \sin \hat{\theta} e^{\hat{\rho}
\sigma_1 \otimes \sigma_3^{\smc{cc}}} \sigma_3 \otimes
\sigma_3^{\smc{cc}}$, where $\hat{\rho}=\diag (\rho_{\smc{bb}} ,
\rho_{\smc{ff}})_{\smc{b,f}}$. In doing so, two coupled equations for
$\hat{\theta}$ and $\hat{\rho}$ are obtained:
\begin{gather}
\label{eq:notai}
   D \nabla^2 \hat{\theta} +
  \Frac{D}{2} \sin 2 \hat{\theta} \left[(\nabla \hat{\rho})^2 +
  4|\vect{h}_0|^2 +4 \vect{h}_0 \cdot \nabla \hat{\rho}\right] + 2 y
  \sin \hat{\theta} =0 \\
  \nabla \cdot \left[\sin^2 \hat{\theta} \left(\nabla \hat{\rho}+ 2
  \vect{h}_0\right) \right] =0 \; .
\notag
\end{gather}
To understand the nature of these equations, let us focus on the quasi
one-dimensional system. In this case the second equation can be
integrated and substituted in the first one, giving respectively
\begin{align*} 
  D \nabla^2 \hat{\theta} + D \vect{j}^2 \Frac{\cos
  \hat{\theta}}{\sin^3 \hat{\theta}} + 2 y \sin \hat{\theta} &=0 &
  \nabla \hat{\rho} &= -2 \vect{h}_0 + \Frac{\vect{j}}{\sin^2
  \hat{\theta}} \; ,
\end{align*}
where $\vect{j}$ is a constant fixed by the boundary conditions. From
these equations one can identify a homogeneous supersymmetric
solution, $\nabla \hat{\rho} =0$ and
\begin{equation}
  \cos \theta_{\smc{mf}} =
  \begin{cases}
  - y \tau_n           & \quad |y| \tau_n < 1 \\
  - \sign (y)          & \quad |y| \tau_n \ge 1 \; ,
  \end{cases}
\label{eq:mfcon}
\end{equation}
where $1/\tau_n = 2D |\vect{h}_0|^2$. Again, making use of
Eq.~\eqref{eq:sourc}, the homogeneous solution translates to a complex
DoS~\eqref{eq:mfdos}, which is flat and non-vanishing only in the
interval $|y| \tau_n <1$. Now, in the imaginary scalar potential
problem, an instanton configuration of the saddle-point equation was
signalled by the existence of tail states. What happens in the present
case? Since, at infinity, the instanton solution is constrained to be
the homogeneous one, the constant $\vect{j}$ is determined by the
homogeneous mean-field solution, $\vect{j}= 2 \vect{h}_0 \sin^2
\theta_{\smc{mf}}$. Therefore, using~\eqref{eq:mfcon}, one can see
that outside the bulk of the spectral support $\vect{j}$
vanishes. Therefore, the equation of motion for $\hat{\theta}$ does
not admit the existence of an instanton. As expected from the
qualitative discussion above, tails are therefore excluded from the
constant imaginary vector potential problem.

Similarly, in the random vector potential system with long-ranged 
correlations (i.e. for $f(|\vect{q}|)=1$), we can expect the paramagnetic 
term to present a similar role. Indeed, subjecting the generating 
functional to an average over the random distribution~\eqref{eq:corre},
one obtains a non-local interaction of the fields. In the present 
context, such terms have been shown by Taras-Semchuk and 
Efetov~\cite{taras_efetov} to reproduce the known renormalisation 
properties of the superdiffusive system.
However, in the present case, we have limited our considerations to 
correlations of the vector field which extend over a maximum range 
$\ell_h$. Then, taking the relevant field configurations of $Q$ (i.e. 
those involving the instanton solution) to be long-ranged on a scale 
$r_0$ greatly in excess of $\ell_h$, the vector field entering the 
paramagnetic term can be replaced by its spatial average, i.e. the 
paramagnetic term can be eliminated. By contrast, the diamagnetic 
contribution survives spatial averaging. The total action assumes the 
form
\begin{equation}
   S [Q] = -\Frac{\pi \nu}{8} \int d \vect{r} \; \str \left[D
  (\nabla Q)^2 - 4 \left(\eta \sigma_3 \otimes
  \sigma_3^{\smc{cc}} - y \sigma_1\right) Q + \Frac{1}{\tau_n}
  \left(Q \sigma_1 \otimes \sigma_3^{\smc{cc}}\right)^2 \right] \; ,
\label{eq:vnlsm2}
\end{equation}
where, setting $\tilde{\gamma}_1 = \gamma_1 L^d \int d
\vect{q}/(2\pi)^d \; f (|\vect{q}|)$,
\begin{equation*}
  \Frac{1}{\tau_n} = 2 (d-1) D \tilde{\gamma}_1 \; .
\end{equation*}

\begin{figure}
\begin{center}
\includegraphics[width=0.6\linewidth,angle=0]{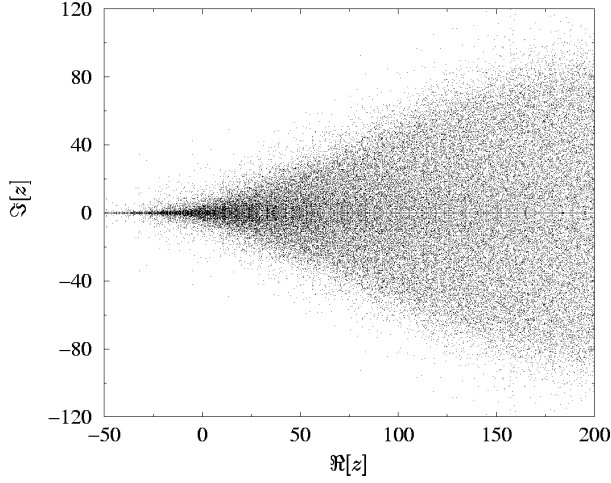}
\end{center}
\caption{\small
        Complex eigenvalues for the for several realizations of the 
        non-Hermitian vector potential Hamiltonian, where the 
        potential $W$ is drawn from a Gaussian 
        $\delta$-correlated random impurity distribution. Here we have 
        included a lattice of $29 \times 29$ $\vect{q}$-points in the
        momentum space. Notice that the constraints on the numerical
        simulation limit the relevant data to the interval
        $x=\Re[z]<200$.}
\label{fig:specv}
\end{figure}

Now, as with the imaginary scalar potential, the profile of the bulk 
DoS and tails states can be obtained by subjecting the action to a
saddle-point analysis. Varying the action with respect to $Q$, subject 
to the non-linear constraint, $Q^2=\mathbb{I}$, one obtains the 
homogeneous saddle-point equation, 
\begin{equation*}
  D\nabla(Q\nabla Q)-[\eta \sigma_3 \otimes
  \sigma_3^{\smc{cc}} - y \sigma_1 , Q] + \Frac{1}{2
  \tau_n} [\sigma_1\otimes\sigma_3^{\smc{cc}} Q \sigma_1 \otimes
  \sigma_3^{\smc{cc}}, Q] = 0 \; ,
\end{equation*}
Then, applying the \emph{Ansatz} $Q = \cos\hat{\theta} \sigma_3 \otimes 
\sigma_3^{\smc{cc}} +\sin \hat{\theta} \sigma_1$, where $\hat{\theta} = 
\diag (i\theta_{\smc{bb}} , \theta_{\smc{ff}})_{\smc{b,f}}$, the saddle-point
equation coincides with Eq.~\eqref{eq:ineqn} allowing the results of 
sections~\ref{sec:state} and~\ref{sec:highd} to be imported. As a result, 
we can immediately deduce the homogeneous mean-field 
solution~\eqref{eq:mfsol} as well as the inhomogeneous instanton 
solution~\eqref{eq:ineqn} of the saddle-point equation. Thus, from the 
homogeneous mean-field solution, one obtains the expression~\eqref{eq:mfdos} 
for the DoS, i.e. the complex DoS is flat and non-vanishing over the interval 
\begin{equation*}
  |y| < \Frac{1}{\tau_n}= \frac{4 (d-1)}{d} \frac{\tilde{\gamma}_1
  |\tau}{m} |x| \; .
\end{equation*}
i.e., in the two-dimensional system, the support for the DoS occupies a
wedge of the complex plane with a width that scales in proportion to
$\tilde{\gamma}_1 |x|$ (see Fig.~\ref{fig:vedos}). Furthermore,
making use of Eqs.~\eqref{eq:fldos} and \eqref{eq:ddima}, one obtains 
complex DoS in the tail region,
\begin{equation*}
  \langle \nu(z) \rangle_{W,h} \sim \exp\left\{-4 \pi a_d
  (\Delta\tau_n)^{(d-2)/2} g^{d/2}
  \left[2(\tau_n|y|-1)\right]^{(4-d)/2}\right\} \; ,
\end{equation*}
where, as usual, $g=\nu D L^{d-2}$ and $\Delta=(\nu L^d)^{-1}$ represent
respectively the dimensionless conductance and average level spacing
of the Hermitian system.

\begin{figure}
\begin{center}
\includegraphics[width=0.6\linewidth,angle=0]{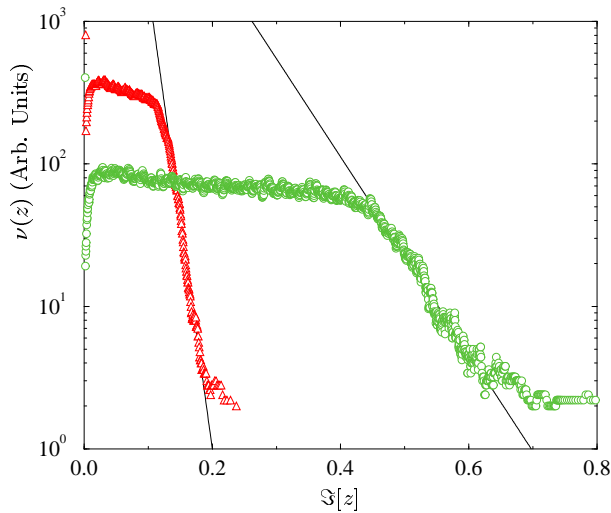}
\end{center}
\caption{\small
        (Logarithm of the) complex DoS versus $\Im [z] = y$ for a
        fixed value of $\Re [z] = x $ taken from the numerical
        simulation of Fig.~\ref{fig:specv} (triangles). A second data
        set (circles) is shown for a value of $\tilde{\gamma}_1$ four
        times as large. Notice that, as expected from the theory, the
        change in $\tilde{\gamma}_1$ is reflected in the slope of the
        exponent. The exponential fits of the two data are shown as
        solid curves.}
\label{fig:fig8a}
\end{figure}

In fact, the behaviour of the vector potential system differs from the 
scalar potential system only in the symmetry of the soft fluctuations
around the saddle-point solution. As usual, the homogeneous saddle-point 
solution $Q_{\smc{mf}}$ is not unique but is spanned by an entire
manifold of solutions $Q = T Q_{\smc{mf}} T^{-1}$ where for $y\ne 0$, 
$[T,\sigma_1] = 0 = [T,\sigma_1 \otimes \sigma_3^{\smc{cc}}]$ with
$T =\gamma (T^{-1})^{\mathsf{T}} \gamma^{\mathsf{T}}$ (Chiral symmetry 
class AIII in the classification of Ref.~\cite{zirnbauer}), while, for $y=0$, 
$[T, \sigma_1 \otimes \sigma_3^{\smc{cc}}] = 0$ (Chiral symmetry class 
BDI). Now while the soft (class AIII) fluctuations decouple from the DoS 
source (c.f. the scalar potential Hamiltonian), the soft (class BDI) 
fluctuations couple. As a result, the complex DoS becomes strongly 
suppressed near $y=0$ while, on the real axis, a finite density of 
eigenvalues accumulates --- see below. 

The amalgamation of data for the eigenvalues of ca. $100$ realizations of 
the imaginary vector potential Hamiltonian is shown in Fig.~\ref{fig:specv}. 
The locus of the edge of the support predicted by the mean-field theory 
is in good agreement with the numerics, while the depletion of eigenvalues
from the interval near $y=0$ and the singular density at $y=0$ are clearly
resolved. In particular, a comparison of the complex DOS with the numerical 
simulation shown in Fig.~\ref{fig:dos_14} shows a good agreement with 
theory.

\begin{figure}
\begin{center}
\includegraphics[width=0.6\linewidth,angle=0]{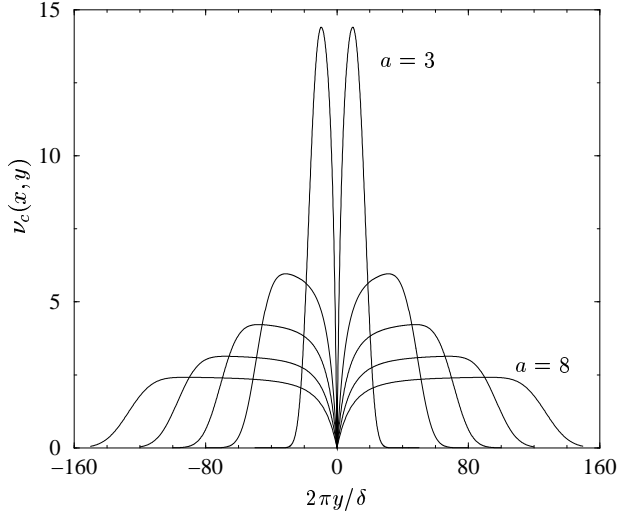}
\end{center}
\caption{\small
        Complex DoS $\nu_c (z)$ as a function of $2 \pi y / \Delta$
        for $\nu = 10$ and $a = 3,5,6,7,8$. 
        }
\label{fig:efeto}
\end{figure}

\subsection{Zero-dimensional Limit}
\label{sec:zerod}
Finally, to conclude our discussion, let us turn to consider the
properties of the imaginary vector potential Hamiltonian when the
system enters the zero-dimensional limit. Our analysis above shows
that, outside the region of support, the spectrum is characterized by
tail states which are localized on a length scale of $r_0
(\kappa)$, \eqref{eq:dropl}. When the size of the droplet region
becomes in excess of the system size (inevitable in a finite system
as one approaches the edge of the support, $\kappa \to 1^+$), the system 
enters a zero-dimensional  
regime where the action is dominated by the zero spatial mode of
$Q$. Here the action collapses onto the zero-dimensional theory
analyzed in detail by Efetov~\cite{efetov1}. 

In this limit, we can proceed in two ways: firstly, following the discussion
above, we can apply the saddle-point analysis seeking a symmetry
broken field configuration. However, in this case, the relative of the
bounce solution is now a stationary point of the zero-dimensional
potential $V(\phi)$. Alternatively, following Efetov~\cite{efetov1}
one can undertake an exact evaluation of the zero-dimensional
$\sigma$-model. A pursuit of the second route obtains the exact
formula for the complex DoS. The latter can be
separated into two contributions $\nu (z) = \nu_r (z) + \nu_c(z)$,
where
\begin{align*}
  \nu_r (x , y) &= \nu \delta(y) \int_0^1 dt \;
  e^{- a^2 t^2} &
  \nu_c (x , y) &= \Frac{\pi \nu}{\Delta} \Phi\left(\Frac{\chi}{2 a}\right)
  \int_0^1 dt \; t \sinh (\chi t) e^{-a^2 t^2} \; ,
\end{align*}
where
\begin{align*}
  \chi &= \Frac{2 \pi |y|}{\Delta} & a^2 = \Frac{\pi}{\Delta\tau_n} \; ,
\end{align*}
with $\Phi (u) = 2 \int_{u}^\infty dt \; e^{-t^2} / \sqrt{\pi}$ and 
$\Delta$ is the level spacing of the Hermitian system. The presence of
the anomalous part $\nu_r (x , y)$ attests that a finite fraction of
eigenvalues remains real. Otherwise the component $\nu_c (x , y)$
describes the distribution of complex eigenvalues and its form is 
shown in Fig.~\ref{fig:efeto}. In the region $\chi
\gg 2a$ the DoS coincides with the constant mean-field
value~\eqref{eq:mfdos} for $\chi < 2a^2$, while outside this interval
one has the following asymptotic expansion:
\begin{equation}
  \nu_c (x , y) \Simiq_{\chi > 2 a^2} \Frac{\sqrt{\pi} \nu}{\Delta}
  \Frac{a}{\chi^2} \exp \left( \chi - \Frac{\chi^2}{4 a^2} - a^2
  \right) \; .
\label{eq:asymp}
\end{equation}

With the exact result at hand, let us compare Eq.~\eqref{eq:asymp} with
the result of the saddle-point analysis from the symmetry broken field
configuration. In particular, in the zero-dimensional case, the 
saddle-point analysis leads to the `instanton' action
\begin{equation*}
  S_{\text{inst}} = \Frac{\pi|y|}{2 \Delta} [ V_{\smc{r}} (0) -
  V_{\smc{r}} (\phi_{\text{min}})] \; ,
\end{equation*}
where $\phi_{\text{min}}$ is the minimum of the potential $V_{\smc{r}}
(\phi)$ (Fig.~\ref{fig:poten}). Evaluating this minimum, one obtains
\begin{equation*}
  S_{\text{inst}} = \Frac{\pi}{\Delta\tau_n} 
  \left(|y|\tau_n - 1 \right)^2 = \Frac{\chi^2}{4
  a^2} - \chi + a^2 \; .
\end{equation*}
From the expression for the DoS~\eqref{eq:fldos}, we can conclude that,
in the zero-dimensional case, the exponential dependence of the DoS
coincides with the exact result obtained in Ref.~\cite{efetov1}. 

\section{Conclusions}
\label{sec:concl}
To conclude, we have implemented a field theoretic scheme to explore
the structure of the DoS close to the edge of the support of two linear
non-Hermitian operators describing a quantum particle subject to a
random imaginary scalar and an imaginary vector potential. In doing so, we 
have provided a general scheme for the symmetry classification of 
non-Hermitian operators. The field theoretic approach is easily generalized
to the consideration of higher point spectral correlations of the 
fields.

In the quasi-classical limit, where the real part of the energy is in 
excess of any other energy scale, the tails are dominated by `optimal 
configurations' of the real random scalar potential. In contrast to band 
tail states in semi-conductors, these states are quasi-classical in nature
being localized on the length scale $\xi=(D/|y|)^{1/2}\gg \ell$. As such
the profile of the DoS and their dimension is universal depending on
just a few material parameters and independent of the nature of the
impurity distribution. In the particular case of the constant imaginary 
vector potential, we have argued that tail states of the system are 
prohibited by the delocalization mechanism of Hatano and Nelson. 

\paragraph{Acknowledgments:} We are grateful to Victor Gurarie and Damian
Taras-Semchuk for useful discussions. One of us (FMM) would like to 
acknowledge the financial support of Scuola Normale Superiore.


\begin{thebibliography}{99}
%

\bibitem{sommers}
H. J. Sommers, A. Crisanti, H. Sampolinski and Y. Stein,
\emph{Phys.~Rev.~Lett.} {\bf 60}, 1895 (1988).

\bibitem{haake}
F. Haake, F. Izrailev, N. Lehmann, D. Saher and H. -J. Sommers,
\emph{Z.~Phys.~B} {\bf 88}, 359 (1992).

\bibitem{hatano_nelson}
N. Hatano and D. R. Nelson, \emph{Phys.~Rev.~Lett.} \textbf{77}, 570
(1996); \emph{Phys.~Rev.~B} {\bf 56}, 8651 (1997).

\bibitem{efetov1}
K. B. Efetov, \emph{Phys.~Rev.~B} \textbf{56}, 9630 (1997);
\emph{Phys.~Rev.~Lett.} \textbf{79}, 491 (1997).

\bibitem{brouwer}
P. W. Brouwer, P. G. Silvestrov and C. W. J. Beenakker,
\emph{Phys.~Rev.~B} {\bf 56}, R4333 (1997).

\bibitem{fyodorov}
Y. V. Fyodorov, B. A. Khoruzhenko and H. -J. Sommers,
\emph{Phys.~Lett.~A} {\bf 226}, 46 (1997); \emph{Phys.~Rev.~Lett.}
{\bf 79}, 557 (1997).

\bibitem{janik}
R. A. Janik, M. A. Nowak, G. Papp and I. Zahed, \emph{Nucl.~Phys.~B}
{\bf 501}, 603 (1997).

\bibitem{feinberg_zee}
J. Feinberg and A. Zee, \emph{Nucl.~Phys. B} \textbf{504}, 579
(1997).

\bibitem{chalker_wang1}
J. T. Chalker and Z. J. Wang, \emph{Phys.~Rev.~Lett.}, \textbf{79},
1797 (1997).

\bibitem{brezin_zee}
E. Br\'ezin and A. Zee, \emph{Nucl.~Phys.~B} \textbf{509}, 599 (1998);
J. Feinberg and A. Zee, \emph{Phys.~Rev.~E} \textbf{59}, 6433 (1999). 

\bibitem{mudry_simons}
C. Mudry, B. D. Simons and A. Altland, \emph{Phys.~Rev.~Lett.} {\bf
80}, 4257 (1998).

\bibitem{chalker_mehlig}
J. T. Chalker and B. Mehlig, \emph{Phys.~Rev.~Lett.} {\bf 81}, 3367
(1998); B. Mehlig and J. T. Chalker, preprint cond-mat/9906279.

\bibitem{hastings} 
M. B. Hastings, preprint cond-mat/9909234.

\bibitem{janik2} 
R. A. Janik, W. Noerenberg, M. A. Nowak, G. Papp and I. Zahed
\emph{Phys.~Rev.~E} {\bf 60}, 2699 (1999).

\bibitem{izyumov_simons1}
A. V. Izyumov and B. D. Simons, \emph{Europhys.~Lett.} \textbf{45},
290 (1999).

\bibitem{izyumov_simons2}
A. V. Izyumov and B. D. Simons, \emph{Phys.~Rev.~Lett.} \textbf{83},
4373 (1999).

\bibitem{yurkevich}
I. V. Yurkevich and I. V. Lerner, \emph{Phys.~Rev.~Lett.} {\bf 82},
5080 (1999).

\bibitem{kolesnikov_efetov}
A. V. Kolesnikov and K. B. Efetov, \emph{Phys.~Rev.~Lett.}
\textbf{84}, 5600 (2000).

\bibitem{mudry_brouwer}
C. Mudry, P. W. Brouwer, B. I. Halperin, V. Gurarie, and A. Zee,
\emph{Phys.~Rev. B} \textbf{58}, 13539 (1998).

\bibitem{edwards}
S. F. Edwards, \emph{Proc.~Phys.~Soc.} {\bf 85}, 613 (1965).

\bibitem{kleinert}
H. Kleinert, \emph{Path Integrals in Quantum Mechanics, Statistics and
Polymer Physics}, World Scientific, Singapore London (1995).

\bibitem{bouchaud_georges}
J. P. Bouchaud and A. Georges, \emph{Phys.~Rep.} {\bf 195}, 127
(1990).

\bibitem{isichenko}
M. B. Isichenko, \emph{Rev.~Mod.~Phys.} {\bf 64}, 961 (1992).

\bibitem{abrahams_anderson}
E. Abrahams, P. W. Anderson, D. C. Licciardello and
T. V. Ramakrishnan, \emph{Phys.~Rev.~Lett.} \textbf{42}, 673 (1979).

\bibitem{gorkov_larkin}
L. P. Gor'kov, A. I. Larkin and D. E. Khmel'nitskii, \emph{JETP~Lett.}
\textbf{30}, 228 (1979).

\bibitem{ginibre} 
J. Ginibre, \emph{J.~Math.~Phys.} {\bf 6}, 440 (1965).

\bibitem{girko}
V. L. Gir'ko, \emph{Theor.~Prob.~Appl.} {\bf 29}, 694 (1985).

\bibitem{balagurov_vaks}
Ya. B. Balagurov and V. G. Vaks, \emph{Sov.~Phys. JETP} {\bf 38}, 968
(1974).

\bibitem{samokhin}
K. Samokhin, \emph{J.~Phys.~A} {\bf 31}, 9455 (1998).

\bibitem{shnerb}
N. M. Shnerb, \emph{Phys.~Rev.~B} {\bf 57}, 8571 (1998).

\bibitem{lifshitz}
I. M. Lifshitz, \emph{Sov.~Phys.~Usp.} {\bf 7}, 549 (1965);
\emph{Adv.~Phys.} {\bf 13}, 483 (1964); \emph{Zh.~Eksp. Teor.~Fiz.}
{\bf 53}, 743 (1967).

\bibitem{zittartz_langer}
J. Zittartz and J. S. Langer, \emph{Phys.~Rev.} {\bf 148}, 741 (1966).

\bibitem{halperin_lax}
B. I. Halperin and M. Lax, \emph{Phys.~Rev.} {\bf 148}, 722 (1966).

\bibitem{gredeskul}
I. M. Lifshitz, S. A. Gredeskul and L. A. Pastur,
\emph{Sov.~J. Low~Temp.~Phys.} {\bf 2}, 533 (1976).

\bibitem{efetov}
K. B. Efetov, \emph{Supersymmetry in Disorder and Chaos}, Cambridge
University Press, Cambridge (1997).

\bibitem{zirnbauer}
M. R. Zirnbauer, \emph{J.~Math.~Phys.} \textbf{37}, 4986 (1996).

\bibitem{altland_simons}
A. Altland, B. D. Simons and D. Taras-Semchuk, \emph{Adv.~Phys.}
\textbf{49}, 321 (1998).

\bibitem{bundschuh}
R. Bundschuh, C. Cassanello, D. Serban and M. R. Zirnbauer,
\emph{Nucl.~Phys.~B} \textbf{532}, 689 (1998).

\bibitem{simons_altland}
B. D. Simons and A. Altland, ``Mesoscopic Physics'', to be published
in the Proceedings of the CRM Summer School \emph{Theoretical Physics
at the End of the XXth Century} (Banff, Canada, 1999), CRM Series in
Mathematical Physics, Springer, Berlin (2000).


\bibitem{abrikosov_gorkov}
A. A. Abrikosov and L. P. Gor'kov, \emph{Sov.~Phys. JETP} \textbf{12},
1243 (1961).

\bibitem{lamacraft_simons1}
A. Lamacraft and B. D. Simons, \emph{Phys.~Rev.~Lett.} \textbf{85},
4783 (2000); preprint cond-mat/0101080.

\bibitem{coleman}
S. Coleman, \emph{Aspects of Symmetry. Selected Erice   
lecture}. Cambridge University Press, Cambridge, (1985).

\bibitem{derrida}
B. Derrida and J. M. Luck, \emph{Phys.~Rev.~B} \textbf{28}, 7183
(1983).

\bibitem{kravtsov}
V. E. Kravtsov, I. V. Lerner and V. I. Yudson, \emph{Sov.~Phys. JETP}
\textbf{64}, 336 (1986).

\bibitem{fisher}
D. S. Fisher, \emph{Phys.~Rev. A} \textbf{30}, 960 (1984).

\bibitem{taras_efetov}
D. Taras-Semchuk and K. B. Efetov, \emph{Phys.~Rev. B} \textbf{64},
115301 (2001).

\end{thebibliography}
\end{document}